\documentclass[10pt,journal,compsoc]{IEEEtran} %

\usepackage{amssymb}
\usepackage{amsmath}
\usepackage{graphicx, subfigure}
\usepackage{tabularx}
\usepackage{multirow}
\usepackage{makecell}
\usepackage{url}
\usepackage{float}
\usepackage{bm}
\usepackage{color}
\usepackage{booktabs}
\usepackage{threeparttable}
\usepackage{xcolor}
\usepackage{adjustbox}
\usepackage[ruled]{algorithm2e} 

\newcommand{\tabincell}[2]{\begin{tabular}{@{}#1@{}}#2\end{tabular}}

\begin{document}

\title{StrucADT: Generating Structure-controlled 3D Point Clouds with Adjacency Diffusion Transformer}

\author{Zhenyu~Shu,
	Jiajun~Shen*,
	Zhongui~Chen,
	Xiaoguang~Han,
	and~Shiqing~Xin 
	\IEEEcompsocitemizethanks{
		\IEEEcompsocthanksitem Zhenyu Shu is with School of Computer and Data Engineering, NingboTech University, Ningbo 315100, China. He is also with Ningbo Institute, Zhejiang University, Ningbo 315100, China (e-mail: shuzhenyu@nit.zju.edu.cn).
		\IEEEcompsocthanksitem Jiajun Shen is with School of Software Technology, Zhejiang University, Ningbo 315048, China (e-mail: jiajunshen\_paper@163.com). Corresponding author.
		\IEEEcompsocthanksitem Zhonggui Chen is with the School of Informatics, Xiamen University, Xiamen 361005, China (e-mail: chenzhonggui@xmu.edu.cn).
		\IEEEcompsocthanksitem Xiaoguang Han is with Shenzhen Research Institute of Big Data, Chinese University
		of Hong Kong, Shenzhen 518172, China. (e-mail: hanxiaoguang@cuhk.edu.cn).		
		\IEEEcompsocthanksitem  Shiqing Xin is with School of Computer Science and Technology, ShanDong University, Jinan 250100, China (e-mail: xinshiqing@sdu.edu.cn).
	}
	\thanks{Manuscript received month day, year; revised month day, year.}
}

\markboth{IEEE transactions on visualization and computer graphics,~Vol.~XX, No.~X, Month~Year}%
{Shell \MakeLowercase{\textit{et al.}}: Bare Demo of IEEEtran.cls for Computer Society Journals}

\IEEEtitleabstractindextext{
\begin{abstract}
	In the field of 3D point cloud generation, numerous 3D generative models have demonstrated the ability to generate diverse and realistic 3D shapes. However, the majority of these approaches struggle to generate controllable 3D point cloud shapes that meet user-specific requirements, hindering the large-scale application of 3D point cloud generation. To address the challenge of lacking control in 3D point cloud generation, we are the first to propose controlling the generation of point clouds by shape structures that comprise part existences and part adjacency relationships. We manually annotate the adjacency relationships between the segmented parts of point cloud shapes, thereby constructing a StructureGraph representation. Based on this StructureGraph representation, we introduce StrucADT, a novel structure-controllable point cloud generation model, which consists of StructureGraphNet module to extract structure-aware latent features, cCNF Prior module to learn the distribution of the latent features controlled by the part adjacency, and Diffusion Transformer module conditioned on the latent features and part adjacency to generate structure-consistent point cloud shapes. Experimental results demonstrate that our structure-controllable 3D point cloud generation method produces high-quality and diverse point cloud shapes, enabling the generation of controllable point clouds based on user-specified shape structures and achieving state-of-the-art performance in controllable point cloud generation on the ShapeNet dataset.
\end{abstract}

}

\maketitle

\IEEEdisplaynontitleabstractindextext

\IEEEpeerreviewmaketitle


\IEEEraisesectionheading{\section{Introduction}\label{sec:introduction}}
\IEEEPARstart{3}D shape generation~\cite{xu2023dl3dsurvey,shi2022dgm3dsurvey} is a fundamental task in computer graphics, holding significant importance across various domains, including modeling, animation, and gaming industries. The field of 3D shape generation has witnessed substantial progress, with numerous state-of-the-art methods and applications emerging in voxels~\cite{girdhar2016tlnet,tatarchenko2017octree,wu20163dgan,chen2019text2shape,zhou2021pvd,mittal2022autosdf}, point clouds~\cite{yang2019pointflow,luo2021dpm,nichol2022pointe,nakayama2023difffacto}, meshes~\cite{groueix2018atlasnet,wang2018pixel2mesh,wen2019pixel2mesh++,nash2020polygen,siddiqui2024meshgpt,alliegro2023polydiff}, and structured representations~\cite{li2017grass,mo2019structurenet,gao2019sdmnet,mo2019partnet,yang2022dsgnet}. 3D point cloud shape data can be easily obtained through 3D scanners, making the generation of 3D point cloud shapes a topic of extensive attention and research.

In the realm of 3D point cloud generation, many 3D generative models, such as PointFlow~\cite{yang2019pointflow}, DPM~\cite{luo2021dpm}, Point-E~\cite{nichol2022pointe}, and DiffFacto~\cite{nakayama2023difffacto}, have shown proficiency in producing diverse and realistic 3D shapes. Methods like PointFlow, DPM, and DiffFacto treat the 3D point cloud generation process as a distribution of distributions~\cite{yang2019pointflow}, first sampling from the overall shape distribution of a class of point clouds and then sampling the distribution of points from the given shape, thus enabling the generation of point clouds with an arbitrary number of points. These methods employ flow-based models~\cite{dinh2017realnvp,kingma2018glow,rezende2015variational,chen2018neural} or diffusion models~\cite{sohl2015deepnt,song2019ncsn,ho2020ddpm,song2021ddim,nichol2021iddpm,dhariwal2021adm,song2021sde,rombach2022ldm} to learn the aforementioned distributions.

\begin{figure}
	\centering
	\includegraphics[width=0.48\textwidth]{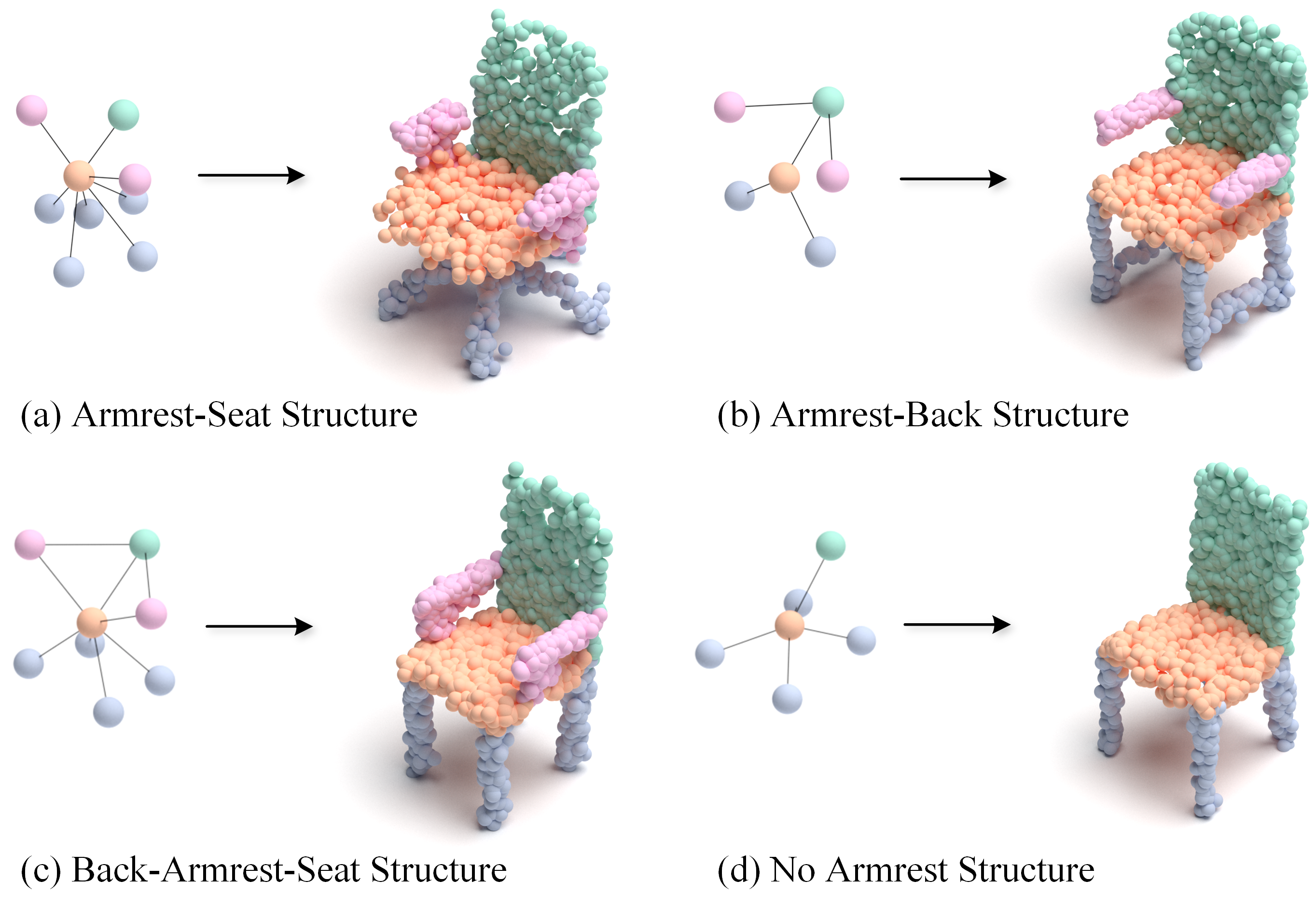}
	\caption{
		(a) The Armrest-Seat structure to control the chair generated with the armrest attached to the seat, (b) the Armrest-Back structure to control the chair generated with the armrest attached to the back, (c) the Back-Armrest-Seat structure to control the chair generated with the armrest attached to the back and the seat, and (d) the No-Armrest structure to control the chair generated without the armrest.
	}
	\label{fig:motivation}
\end{figure}

However, most of these approaches struggle to generate controllable 3D point cloud shapes. These methods can only learn to approximate the shape distribution from the dataset and sample point cloud shapes from the learned distribution, but they lack the ability to control the generation of shapes that meet user-specific requirements, as users usually want to generate shapes that fit their needs. Some multimodal generation methods, such as Point-E~\cite{nichol2022pointe}, can control the generation process with text or image conditions. However, these methods require text or image data matching the point cloud shapes, thus necessitating extensive manual annotation, which is costly and challenging.

In order to address the challenge of lacking control in 3D point cloud generation, we explore the inherent structure of 3D shapes. The structure of a shape is whether the parts that make up the shape appear and whether these parts are adjacent. For shapes within the same category, although their overall geometric profile may vary significantly, their structures remain similar~\cite{chaudhuri2020lgmstruc}. For example, a chair is typically composed of the back, seat, legs, and armrests, where the seat is usually adjacent to the back and legs. The core observation of this paper is leveraging the structure of 3D point cloud shapes to better control their generation. Figure~\ref{fig:motivation} shows that different structures of the chair are able to control the generation of chairs with different styles: a chair generated with armrests attached to the seat or the back or both the seat and the back, or even without the armrests. By introducing the shape structure, it becomes possible to control the generation of 3D point cloud shapes explicitly and more precisely.

However, existing 3D shape segmentation datasets~\cite{mo2019partnet,chen2009psb,wang2012coseg,maron2017humanbody,chang2015shapenet} only provide semantic segmentation labels without the connectivity information between parts. Although StructureNet~\cite{mo2019structurenet} represents shapes as $n$-ary part hierarchies with adjacency between sibling parts, they simply compute the minimum distance between two segmented parts to obtain the part adjacency on the PartNet~\cite{mo2019partnet} dataset. In contrast, we manually annotate the adjacency relationships between segmented parts on the widely used ShapeNet dataset, forming \textit{StructureGraph} representation for point cloud shapes to control their generation more precisely. In this graph, nodes represent the segmented parts of a point cloud shape, while edges denote the connections between these segmented parts.

To control the 3D point cloud generation process with our \textit{StructureGraph} representation, we introduce \textbf{StrucADT}: Generating \textbf{Struc}ture-controlled 3D point clouds with \textbf{A}djacency \textbf{D}iffusion \textbf{T}ransformer. StrucADT explicitly uses the StructureGraph of 3D shapes as user-specific input to control the 3D point cloud generation. To enable the network to incorporate the feature information from the StructureGraph, we design a \textit{StructureGraphNet} encoder to fuse the structural information between parts. The learned features of the StructureGraph and the part adjacency are then incorporated into our proposed \textit{cCNF Prior} module and \textit{Diffusion Transformer}~\cite{vaswani2017transformer,peebles2023dit} module to generate structure-consistent point cloud shapes.

Our contributions are as follows:
\begin{itemize}
	\item
		We introduce StrucADT: Generating Structure-controlled 3D point clouds with Adjacency Diffusion Transformer, which can produce novel and realistic point cloud shapes controlled by user-specific shape structures. To the best of our knowledge, we are the first to control the 3D point cloud generation with shape structures. 
	\item
		To better control the generation of 3D point clouds using the structure of 3D point cloud shapes, we annotate the adjacency relationships between segmented parts on the ShapeNet dataset, forming \textit{StructureGraph} representation for 3D shapes.
	\item
		We propose \textit{StructureGraphNet} module to extract structure-aware latent features, \textit{cCNF Prior} module to learn the distribution of the latent features controlled by the part adjacency, and \textit{Diffusion Transformer} module conditioned on the latent features and part adjacency to generate structure-controllable point cloud shapes. Extensive experimental results on the ShapeNet dataset showcase that our method achieves state-of-the-art performance in controllable point cloud generation.
\end{itemize}

The rest of the paper is structured as follows. Section~\ref{sec:related_work} provides a review of the related work. Section~\ref{sec:our_method} presents a detailed explanation of our proposed method. In Section~\ref{sec:experiments}, we evaluate the performance of our algorithm on the ShapeNet dataset. Section~\ref{sec:limitation_future_work} discusses the limitations of our approach and suggests future research directions. Lastly, Section~\ref{sec:conclusion} concludes the paper.

\section{Related Work}\label{sec:related_work}

\subsection{3D Point Cloud Generation}

Achlioptas et al.~\cite{achlioptas2018rgan} propose a deep autoencoder network with good reconstruction quality and generalization ability. The network includes an r-GAN running on the original point cloud, an l-GAN with significant improvements trained in the fixed latent space of the autoencoder, and the Gaussian Mixture Models (GMMs). Valsesia et al.~\cite{valsesia2018gcrgan} improve the r-GAN using a generative adversarial network architecture based on graph convolutional networks. This method can dynamically construct a proximity graph during the generation process, enabling the extraction and representation of local features. TreeGAN~\cite{shu2019treegan} enhances feature representation capabilities by performing graph convolutions in a tree, utilizing ancestor information to generate multi-category 3D point clouds in an unsupervised manner.

Most of the above-mentioned 3D point cloud generation models can only generate a fixed number of point clouds. To address this issue, PointFlow~\cite{yang2019pointflow} views the 3D point cloud generation process as sampling a shape from a shape distribution and then sampling a point cloud from the distribution of that shape. The method uses two CNF~\cite{chen2018neural} generation models to model the above process, enabling the generation of point clouds with an arbitrary number of points. Based on PointFlow, DPM~\cite{luo2021dpm} proposes a 3D point cloud generation method built up on diffusion models that model the reverse diffusion process of point clouds as a Markov chain conditioned on shape latent variables.

However, most of these approaches struggle to generate controllable 3D point cloud shapes. To address the challenge of lacking control in 3D point cloud shape generation, we introduce StructureGraph representation and propose a novel part adjacency conditioned Diffusion Transformer module to control the process of 3D point cloud generation.

\subsection{Structure-Aware 3D Shape Generation}
The structure of a shape generally refers to the components that constitute a shape and the connections between these components, which is a high-level abstraction of 3D shapes. In order to enable users to modify or manipulate the synthesized 3D shapes easily, the generative models should be structure-aware, and users should be able to manipulate the generated 3D shapes by modifying the high-level structure of the shapes~\cite{chaudhuri2020lgmstruc}.

Many works represent the structure of shapes as trees or graphs. GRASS~\cite{li2017grass} proposes a new neural network architecture for encoding and synthesizing 3D shapes, particularly their structures. The method employs a recursive neural networks (RvNNs) based autoencoder to map flat, unlabeled, arbitrary parts to compact codes. The code effectively captures the hierarchy of man-made 3D objects with varying structural complexity. Ren et al.~\cite{ren2017jgl} introduce a method for creating consistent visualizations of collections of segmented meshes by embedding graphs jointly, using spectral graph drawing for initialization and stress majorization for refinement, enabling distance preservation and alignment even with partial or soft node correspondences. StructureNet~\cite{mo2019structurenet} is a hierarchical graph network designed to encode and generate diverse, realistic 3D shapes by representing them as $n$-ary graphs, effectively handling continuous deformations and structural alterations. SDM-Net~\cite{gao2019sdmnet} proposes a deep generative neural network for producing structured deformable meshes. The architecture of SDM-Net is a two-level variational autoencoder, ensuring consistency between the overall shape structure and surface details. DSG-Net~\cite{yang2022dsgnet} employs a deep neural network that enhances 3D shape generation by learning a disentangled structured and geometric mesh representation, allowing for sophisticated control over shape synthesis with separate yet synergistic encoding of geometry and structure.

Recent works also focus on 3D shape generation and manipulation using implicit representations, emphasizing part-level approaches to enable precise and flexible editing of shapes. SPAGHETTI~\cite{hertz2022spaghetti} develops a novel generative framework designed for editing and manipulating 3D shapes represented as neural implicit fields. The architecture disentangles shape representations into intrinsic and extrinsic components, facilitating precise part-level control. SALAD~\cite{koo2023salad} proposes a novel 3D generative model designed for high-quality 3D shape generation and manipulation. It employs a cascaded diffusion framework based on part-level implicit representations, enabling both realistic shape generation and intuitive part-level editing without additional training. 3DShape2VecSet~\cite{zhang20233dshape2vecset} introduces a novel neural field-based representation for 3D shape generation, optimized for generative diffusion models. It encodes 3D shapes into a compact latent space using a set of latent vectors and processes them with cross-attention and self-attention mechanisms.

Although most of these structure-aware methods can generate 3D shapes with plausible structures that are easy to manipulate and modify, their generation process can not be controlled by user-specific input. In contrast, our method can generate realistic and diverse point clouds controlled by the input shape structures.

\subsection{Controllable 3D Shape Generation}
Many recent works control the generation of 3D shapes by introducing additional conditions such as text and images.

PSGN~\cite{fan2017psgn} addresses the issue of non-uniqueness of the true 3D point cloud corresponding to a single image by proposing a novel conditional shape sampler. This sampler is capable of predicting multiple plausible 3D point clouds from an input image. Point-E~\cite{nichol2022pointe} proposes two diffusion models to generate point clouds from complex prompts. It first leverages a pretrained text-to-image diffusion model to generate a single synthetic view from the input text and then utilizes a point cloud diffusion conditioned on the synthetic image to produce 3D point clouds. Spice-E~\cite{sella2024spice} introduces a shape editing method to edit shapes semantically and transform primitive-based abstractions into highly expressive shapes. It adds structural guidance to 3D diffusion models, extending its usage beyond text-conditional generation. This approach supports a variety of applications, including 3D stylization, semantic shape editing, and text-conditional abstraction-to-3D. CLAY~\cite{zhang2024clay} employs a large-scale generative model designed for creating high-quality 3D assets with controllable features. It supports diverse input modalities, including text, images, point clouds, and voxels, enabling users to create intricate 3D models with minimal expertise. Cheng et al.~\cite{cheng2024learninglayout} propose a generative model based on transformer architecture that creates spatially coherent and user-controllable 2D land-use layouts for virtual worlds by learning from real-world geographic data while incorporating geometric and planning objectives to enhance layout quality and adherence to user controls. DepthGAN~\cite{li2024depthgan} is a novel GAN-based method designed for generating accurate and geometrically consistent depth maps from 2D semantic layouts using semantically aware transformer blocks and a semantically weighted depth synthesis module, enabling controllable and effective 3D scene generation for applications like AR and VR.

DiffFacto~\cite{nakayama2023difffacto} is proposed for controllable part-based 3D point cloud generation. It achieves part-level control over shapes by learning the distribution of shapes through a probabilistic generative model. The core innovation of DiffFacto lies in its ability to generate novel shapes by understanding and manipulating the factorized representation of shapes (i.e., decomposing shapes into independent parts and their configurations). This process is realized through a cross-diffusion network that simulates the complex interactions between different parts of shapes, enabling the generation of coherent and plausible 3D structures under various configurations.

Unlike DiffFacto, which controls the 3D point cloud generation through semantic segmentation labels and part transformations, we introduce StructureGraph representation that consists of part existences and part adjacency relationships, explicitly using the shape structures as user control. To enable the network to incorporate the feature information from the StructureGraph, we design the StructureGraphNet module to fuse the structural information between parts. The learned features of the StructureGraph and the part adjacency are then incorporated into the cCNF Prior module and Diffusion Transformer module to control the generation of corresponding point cloud shapes.

\begin{figure*}[t]
	\centering
	\includegraphics[width=0.90\textwidth]{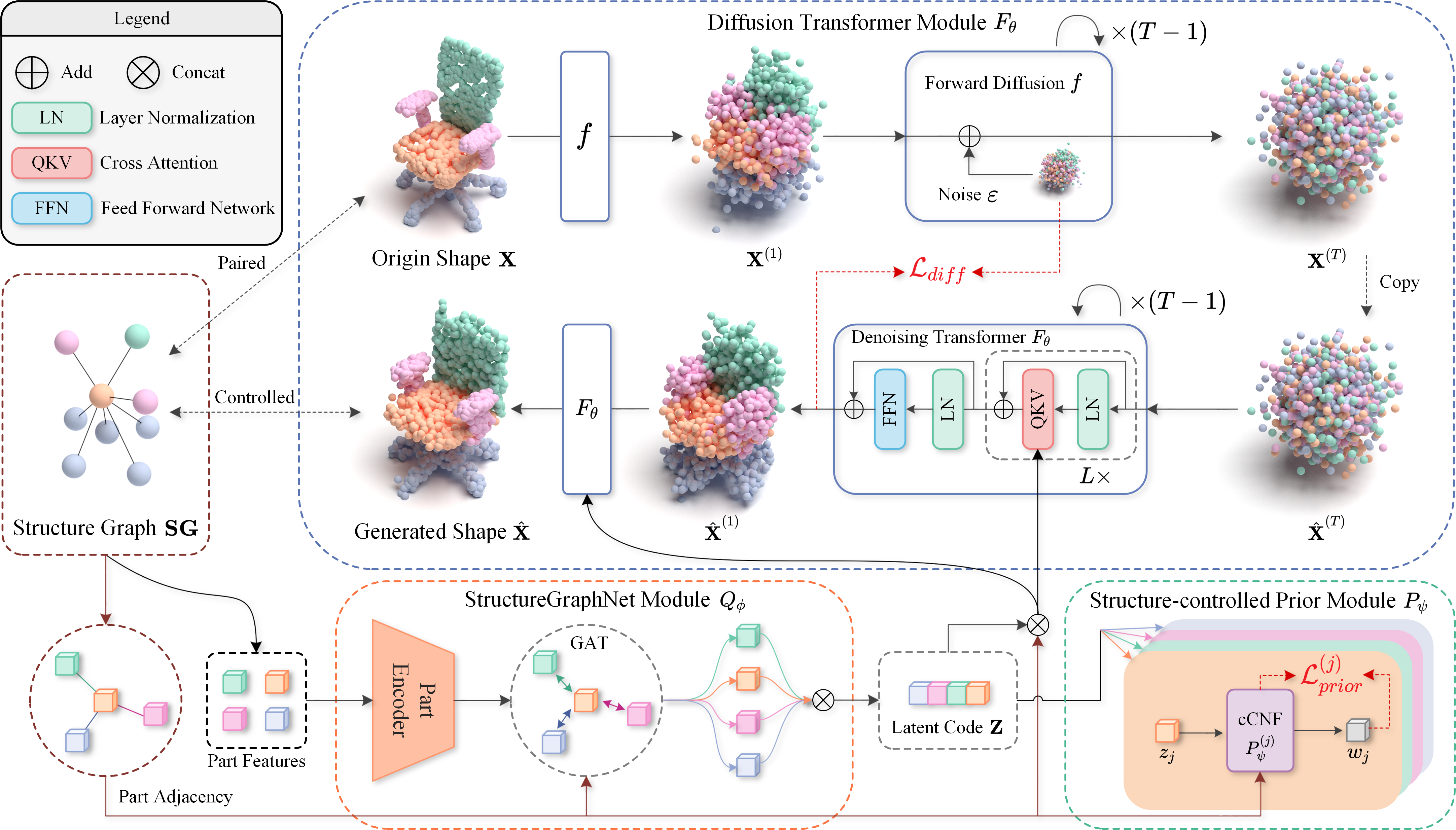}
	\caption{
		Our structure-controlled 3D point cloud generation framework comprises three components: i) The StructureGraphNet Module: This module encodes part features and part adjacency of the StructureGraph into global latent code $ \mathbf{Z} $. ii) The Structure-controlled Prior Module: This module utilizes cCNF models to learn the representation of the global latent code $ \mathbf{Z} $ conditioned on the part adjacency of the StructureGraph. iii) The Diffusion Transformer Module: This module adds noise to 3D point cloud shapes and trains a denoising Transformer controlled by the concatenation of the global label code $ \mathbf{Z} $ and the part adjacency of the StructureGraph. In particular, the part features are the segmentation labels $ \mathbf{S} $ combined with the point cloud shape $ \mathbf{X} $, which is only included in StructureGraph when training. The part adjacency in the figure consists of adjacency relationships $ \mathbf{E} $ and part existences $ \mathbf{V} $ for brevity.
	}
	\label{fig:pipeline}
\end{figure*}

\section{Our method}\label{sec:our_method}
This section begins with an overview of the algorithm's pipeline, followed by individual introductions to each module within the framework. Finally, the complete training and sampling process is presented.

\subsection{Overview}

Each input point cloud shape $ \mathbf{X}=\{x_i\}_{i=1}^n $ consists of $ n $ points, where each point $ x_i \in \mathbb{R}^3 $ represents the three-dimensional coordinates of the point. Each point cloud shape $ \mathbf{X} $ of the same category is segmented into $ m $ parts, forming its semantic segmentation one-hot labels $ \mathbf{S}=\{\{s_{i,j}\}_{j=1}^m\}_{i=1}^n $, where $ s_{i,j} $ represents whether point $ x_i $ is assigned to the $ j $-th part. Based on the segmentation labels $ \mathbf{S} $, we introduce the StructureGraph representation, which consists of part existences $ \mathbf{V}=\{v_j\}_{j=1}^m $ to denote whether part $ j $ exists in shape $ \mathbf{X} $, and adjacency relationships $ \mathbf{E}=\{\{e_{j,k}\}_{k=1}^m\}_{j=1}^m $ to represent whether part $ j $ is adjacent to part $ k $, forming the StructureGraph $ \mathbf{SG}=\{\mathbf{S}, \mathbf{V}, \mathbf{E}\} $. During training, the segmentation labels $ \mathbf{S} $ in $ \mathbf{SG} $ is also combined with the point cloud $ \mathbf{X} $ as part features.

As shown in Figure~\ref{fig:pipeline}, the overall pipeline of our method is divided into the following three modules: i) StructureGraphNet $ Q_\phi $: To extract structure-aware point cloud features $ \mathbf{Z} $. ii) Structure-controlled cCNF Prior Module $ P_\psi $: To learn the distribution of the extracted point cloud features $ \mathbf{Z} $ conditioned on the part adjacency. iii) Structure-controlled Diffusion Transformer Module $ F_\theta $: Uses the structure-aware point cloud features $ \mathbf{Z} $ extracted in i) and part adjacency as conditional context to control the generation of point cloud shapes. 

Note that in Figure~\ref{fig:pipeline}, the part features are the segmentation labels $ \mathbf{S} $ combined with the point cloud shape $ \mathbf{X} $, which is only included in StructureGraph when training. The part adjacency in the figure consists of adjacency relationships $ \mathbf{E} $ and part existences $ \mathbf{V} $ for brevity.

\subsection{StructureGraphNet Module}
We propose a novel StructureGraphNet~(SGN) module to aggregate features of point cloud $ \mathbf{X} $ and StructureGraph $ \mathbf{SG} $, thereby extracting structure-aware latent features $ \mathbf{Z} $. This enables subsequent generative models to generate point clouds based on the latent features and the corresponding structures.

\begin{figure*}[t]
	\centering
	\includegraphics[width=0.90\textwidth]{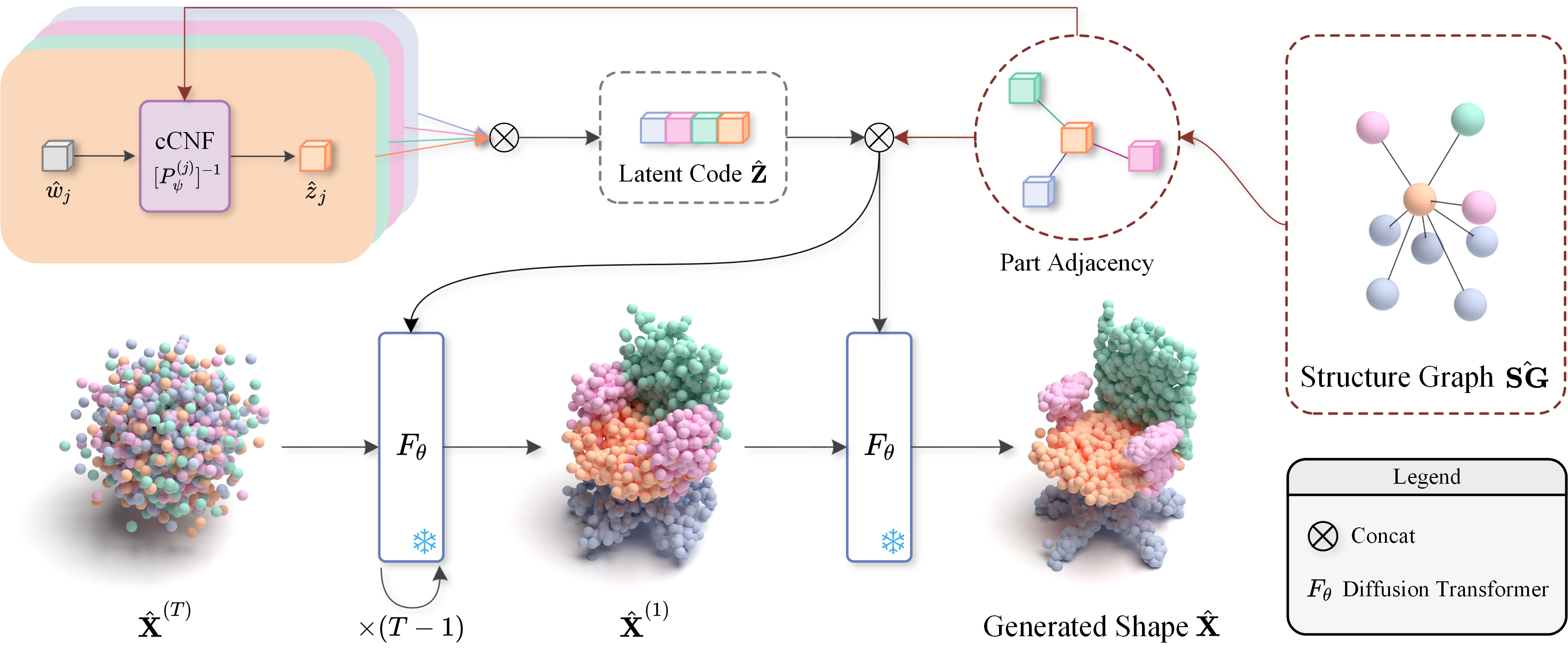}
	\caption{
		Sampling process of our method. For each semantic part, the sampled noise $ \hat w_j $ is converted to $ \hat z_j $ through reversed cCNF, and all are combined to latent code $ \hat{\mathbf{Z}} $. The user-specific StructureGraph $ \hat{\mathbf{SG}} $ contains part adjacency consisting of part existences and adjacency relationships, which are then concatenated with latent code to control the diffusion Transformer $ F_\theta $ gradually sampling novel and controllable point cloud $ \hat{\mathbf{X}} $.
	}
	\label{fig:sampling}
\end{figure*}

The input of the StructureGraphNet is part features and part adjacency. As mentioned above, part features are the combination of the point cloud shape $ \mathbf{X} $ and the segmentation labels $ \mathbf{S} $. For each point cloud shape $ \mathbf{X} $, the StructureGraphNet $ Q_\phi(\mathbf{Z} | \mathbf{X},\mathbf{S},\mathbf{V},\mathbf{E}) $ extracts the structure-aware latent features $ \mathbf{Z}=\{z_j\}_{j=1}^m $:
\begin{equation}\label{eq:strucgn}
	\mathbf{Z} = F_{\mathrm{gat}}\left(F_{\mathrm{conv}}\left(\mathbf{X}\right)^T \cdot \mathbf{S} ,\mathbf{V},\mathbf{E}\right),
\end{equation}
where $ F_{\mathrm{conv}} $ represents a 1D convolutional network to extract the overall feature vector $ \mathbf{Z_g} $ of $ \mathbf{X} $. The semantic segmentation labels $ \mathbf{S} $ of $ \mathbf{X} $ are multiplied by the transposed overall point cloud features $ \mathbf{Z_g}^T $ to extract the local features $ \mathbf{Z_l} $ of each segmented part. Each segmented part $ j $ is treated as a node in the graph, with the extracted local features $ \mathbf{Z_l} $ as the node features. The part existences $ \mathbf{V} $ are utilized as the part mask to ignore the node that does not exist in the graph. The adjacency relationships $ \mathbf{E} $ between parts are considered as edges of the graph. A graph attention network (GAT)~\cite{velivckovic2018gat} $ F_{\mathrm{gat}} $ is applied on this StructureGraph $ \mathbf{SG} $, allowing the network to dynamically adjust the weights between individual parts and other adjacent parts, which enables increased attention to certain part relationships, such as between the armrests and seat in a chair. The final feature vector $ \mathbf{Z} $ is obtained by applying the $ F_{\mathrm{gat}} $ network on the StructureGraph $ \mathbf{SG} $, which incorporates the structural information of the shape parts.

Additionally, to efficiently optimize the Evidence Lower Bound (ELBO), we use reparameterization to sample from the distribution of $ Q_\phi(\mathbf{Z}|\mathbf{X},\mathbf{S},\mathbf{V},\mathbf{E}) $:
\begin{equation}\label{eq:reparameterization}
	\mathbf{Z} =  \mu_{\phi}\left(\mathbf{X},\mathbf{S},\mathbf{V},\mathbf{E}\right) + \sigma_{\phi}\left(\mathbf{X},\mathbf{S},\mathbf{V},\mathbf{E}\right) * \epsilon, 
\end{equation}
where $ \epsilon \sim \mathcal{N}(0,\mathbf{I}) $, $ \mu_{\phi} $ and $ \sigma_{\phi} $ represent the mean and standard deviation of the distribution $ Q_\phi(\mathbf{Z}|\mathbf{X},\mathbf{S},\mathbf{V},\mathbf{E}) $, respectively.

\subsection{Structure-controlled Prior Module}
In the Structure-controlled Prior Module, we employ a conditional Continuous Normalizing Flow (cCNF)~\cite{rezende2015variational,chen2018neural} network to learn the latent features $ \mathbf{Z} $ conditioned on the part existences $ \mathbf{V} $ and adjacency relationships $ \mathbf{E} $. Normalizing flows are a series of reversible mappings that can transform an initial known distribution into a more complex one~\cite{yang2019pointflow}. 


For each point cloud shape $ \mathbf{X} $, the StructureGraphNet $ Q_\phi(\mathbf{Z}|\mathbf{X},\mathbf{S}, \mathbf{V},\mathbf{E}) $ above extracts structure-aware feature vectors $ \mathbf{Z}=\{z_j\}_{j=1}^m $. The structure-controlled cCNF Prior model $ P_{\psi}\left(\mathbf{Z},\mathbf{V},\mathbf{E}\right) = \{P_\psi^{(j)}(z_j,v_j,e_{j.})\}_{j=1}^m $ applies a normalizing flow conditioned on the part adjacency to regularize each part feature vector $ z_j $, yielding a transformed distribution $ \mathbf{W}=\{w_j\}_{j=1}^m $:
\begin{equation}\label{eq:cnf_forward}
	w_j = P_{\psi}^{(j)} \left( z_j,v_j,e_{j.} \right),
\end{equation}
where $ e_{j.} $ represents all the edges adjacent to the part $ j $. 

In the sampling process, as shown in Figure~\ref{fig:sampling}, by inverting this normalizing flow denoted as $ P_{\psi}^{-1} $ and sampling $ \hat w_j $ from a standard Gaussian distribution $ \mathcal{N}(0,\mathbf{I}) $, the distribution of $ z_j $ can be inversely reconstructed conditioned on the user-specific part adjacency $ \hat{v}_j $ and $ \hat{e}_{j.} $:
\begin{equation}\label{eq:cnf_reverse}
	\hat z_j = [P_{\psi}^{(j)}]^{-1} \left(\hat w_j,\hat v_j,\hat e_{j.} \right)
\end{equation}
where $ \hat z_j $ represents the reconstructed distribution of $ z_j $ and $ \hat w_j \sim \mathcal{N}\left(0,\mathbf{I}\right) $ .

Consequently, the loss function for the cCNF Prior can be derived as follows:
\begin{align}\label{eq:cnf_loss}
	\mathcal{L}_{\mathrm{prior}} & = \sum_{j=1}^{m}D_{\mathrm{KL}}\left(Q_{\phi}\left(z_j|\mathbf{X},\mathbf{S},v_j,e_{j.}\right)\left|\right|P_{\psi}^{(j)}\left(z_j,v_j,e_{j.}\right)\right) \notag \\
	                    & = -\sum_{j=1}^{m}\Big[\mathbb{E}_{Q_{\phi}\left(z_j|\mathbf{X},\mathbf{S},v_j,e_{j.}\right)}[\mathrm{log}P_{\psi}^{(j)}\left(z_j,v_j,e_{j.}\right)] \notag \\ 
	                    & + H\left[Q_{\phi}\left(z_j|\mathbf{X},\mathbf{S},v_j,e_{j.}\right)\right]\Big] ,
\end{align}
where $ D_{\mathrm{KL}} $ represents the KL divergence between two distributions and $ H $ is the entropy.

\subsection{Structure-controlled Diffusion Transformer Module}
In the Structure-controlled Diffusion Transformer Module, we leverage the Denoising Diffusion Probabilistic Model (DDPM)~\cite{ho2020ddpm} to learn the conditional likelihood $ P \left(\mathbf{X}|\mathbf{Z},\mathbf{S},\mathbf{V},\mathbf{E}\right)$ through an iterative denoising process.

In the forward diffusion process $ f $, for a point cloud shape $ \mathbf{X} $, Gaussian noise $ \varepsilon $ with $ T $ time steps is added, such that the distribution of the point cloud gradually becomes an independent Gaussian distribution $ f(\mathbf{X}^{(t)}|\mathbf{X}^{(0)}) $:
\begin{align}\label{eq:diff_forward}
	\mathbf{X}^{(t)}                     & = \sqrt{\bar \alpha_t}\mathbf{X}^{(0)} + \sqrt{1-\bar \alpha_t} * \varepsilon \notag \\
	f(\mathbf{X}^{(t)}|\mathbf{X}^{(0)}) & = \mathcal{N}(\mathbf{X}^{(t)}|\sqrt{\bar \alpha_t}\mathbf{X}^{(0)}, \left(1-\bar \alpha_t\right)\mathbf{I}) ,
\end{align}
where $ \varepsilon \sim \mathcal{N}(0, \mathbf{I}) $, $ t=1,\dots,T $, $ \alpha_t=1-\beta_t $, $ \beta_t $ represents the variance schedule that increases with $ t $, $ \bar \alpha_t=\prod_{i=1}^t \alpha_i $.

\begin{algorithm}[tb]\label{alg:training_process}
	\caption{Training process of our method}
	\SetAlgoNoLine

	\KwIn{\\ \quad Point cloud dataset $ \mathbf{X} $ with semantic segmentation labels $ \mathbf{S} $, part existences $ \mathbf{V} $ and part adjacency relationships $ \mathbf{E} $.}
	\KwOut{\\ \quad Trained network parameters $ \phi $ of the SGN $ Q $, $ \psi $ of the Prior model $ P $, and $ \theta $ of the diffusion model $ F $.}
	\textbf{Training process:}\\

	\textbf{Step 1:} Training StructureGraphNet;\\
	\For{$ j \leftarrow 1 $ \KwTo $ m $ }{
		$ Q_\phi \left(z_j|\mathbf{X},\mathbf{S},v_j,e_{j.}\right) \leftarrow \mathbf{X},\mathbf{S},v_j,e_{j.} $\;
		Sample $ \epsilon_j \sim \mathcal{N}(0, \mathbf{I}) $\;
		$ z_j \leftarrow \mu_{\phi}(\mathbf{X},\mathbf{S},v_j,e_{j.}) + \sigma_{\phi}(\mathbf{X},\mathbf{S},v_j,e_{j.}) * \epsilon_j $
	}

	\textbf{Step 2:} Training Prior model;\\
	\For{$ j \leftarrow 1 $ \KwTo $ m $}{
		$ P_\psi^{(j)}(z_j,v_j,e_{j.}) \leftarrow z_j,v_j,e_{j.}  $\;
		\small $ \mathcal{L}_{\mathrm{prior}} \leftarrow D_{\mathrm{KL}}\left(Q_{\phi}\left(z_j|\mathbf{X},\mathbf{S},v_j,e_{j.}\right)\left|\right|P_{\psi}^{(j)}\left(z_j,v_j,e_{j.}\right)\right) $.
	}

	\textbf{Step 3:} Training diffusion model.\\
	\For{$ i \leftarrow 1 $ \KwTo $ n $}{
		Sample $ t \sim \mathrm{Uniform}\{1,\dots,T\} $\;
		Sample $ \varepsilon \sim \mathcal{N}(0, \mathbf{I}) $\; 
		$ x_i^{(0)} \leftarrow x_i $\; 
		$ x_i^{(t)} \leftarrow \sqrt{\bar \alpha_t}x_i^{(0)} + \sqrt{1-\bar \alpha_t} * \varepsilon $ (Eq. (\ref{eq:diff_forward}))\;
		\small $ \mathcal{L}_{\mathrm{diff}} \leftarrow \mathbb{E}_{\varepsilon,t,\mathbf{Z},\mathbf{S},\mathbf{V},\mathbf{E}}\left[\left\|\varepsilon-F_{\theta}(x_i^{(t)},\mathbf{Z},\mathbf{S},\mathbf{V},\mathbf{E})\right\|_2^2\right] $.
	}
\end{algorithm}

\begin{algorithm}[tb]\label{alg:sampling_process}
	\caption{Sampling process of our method}
	\SetAlgoNoLine

	\KwIn{\\ \quad User-specific StructureGraph $ \hat{\mathbf{SG}} $, which consists of segmentation labels $ \hat{\mathbf{S}} $~(can be default), part existences $ \hat{\mathbf{V}} $ and part adjacency relationships $ \hat{\mathbf{E}} $.}
	\KwOut{\\ \quad Generated novel point cloud shapes $ \hat{\mathbf{X}} $ with arbitrary $ \hat{n} $ points.}
	\textbf{Sampling process:}\\

	\textbf{Step 1:} Sampling from the Prior model;\\
	\For{$ j \leftarrow 1 $ \KwTo $ m $}{
		Sample noise $ \hat w_j \sim \mathcal{N}(0,\mathbf{I}) $\;
		$ \hat z_j \leftarrow [P_{\psi}^{(j)}]^{-1} \left(\hat w_j,\hat v_j,\hat e_{j.} \right) $ (Eq. (\ref{eq:cnf_reverse})).
	}

	\textbf{Step 2:} Sampling from the diffusion model;\\
	Sample $ \hat{\mathbf{X}}^{(T)} \sim \mathcal{N}(0,\mathbf{I}) $\;
	\For{$ t \leftarrow T $ \KwTo $ 1 $ }{
		Sample noise $ \epsilon_t \sim \mathcal{N}(0,\mathbf{I}) $\;
		\For{$ i \leftarrow 1 $ \KwTo $ \hat{n} $}{
			$ \mu_0 \leftarrow \frac{1}{\sqrt{\alpha_t}}\hat{x}_i^{(t)} $\;
			$ \mu_1 \leftarrow \frac{1-\alpha_t}{\sqrt{\alpha_t}\sqrt{1-\bar \alpha_t}}F_\theta(\hat{x}_i^{(t)},\hat{\mathbf{Z}},\hat{\mathbf{S}},\hat{\mathbf{V}},\hat{\mathbf{E}}) $\;
			$ \sigma \leftarrow \sqrt{\frac{1-\bar \alpha_{t-1}}{1-\bar \alpha_t}\beta_t} $\;
			$ \hat{x}_i^{(t-1)} = \mu_0-\mu_1+ \sigma * \epsilon_t $ (Eq. (\ref{eq:diff_sampling})).
		}
	}
	$ \hat{\mathbf{X}} \leftarrow \hat{\mathbf{X}}^{(0)} $.
\end{algorithm}

During the reverse diffusion process, the neural network $ F_\theta $ is trained to model the probability distribution, using the forward process $ f(\mathbf{X}^{(t-1)}|\mathbf{X}^{(t)},\mathbf{X}^{(0)}) $ as an approximate posterior to approximately maximize the likelihood $ \mathbb{E}_{f(\mathbf{X}^{(0)})}\mathrm{log}F_\theta(\mathbf{X}^{(0)}) $. Based on the ELBO formula, the simplified DDPM loss function can be derived as the distance between the noise predicted by the network $ F_\theta $ in the reverse diffusion and the ground truth noise $ \varepsilon $ in the forward diffusion process $ f $:
\begin{equation}\label{eq:diff_loss}
	\mathcal{L}_{\mathrm{diff}} = \frac{1}{n}\sum_{i=1}^n\mathbb{E}_{\varepsilon,t,\mathbf{Z},\mathbf{S},\mathbf{V},\mathbf{E}}\left[\left\|\varepsilon-F_{\theta}(x_i^{(t)},\mathbf{Z},\mathbf{S},\mathbf{V},\mathbf{E})\right\|_2^2\right],
\end{equation}
where $ t \sim \mathrm{Uniform}\{1,\dots,T\} $, $ \varepsilon \sim \mathcal{N}(0, \mathbf{I}) $, and $ \mathbf{Z} \sim Q_{\phi} \left( \mathbf{Z}|\mathbf{X},\mathbf{S},\mathbf{V},\mathbf{E} \right) $. The predicted noise $ F_{\theta}(x_i^{(t)},\mathbf{Z},\mathbf{S},\mathbf{V},\mathbf{E}) $ is conditioned on the part features $ \mathbf{Z} $, segmentation semantic labels $ \mathbf{S} $, part existences $ \mathbf{V} $, part adjacency relationships $ \mathbf{E} $, and time step $ t $. Therefore, the denoising process can be controlled through shape structure when sampling.

As shown in Figure~\ref{fig:pipeline}, $ F_\theta $ is a Transformer~\cite{vaswani2017transformer,peebles2023dit} that has $ L $ cross-attention layers and a feedforward network~(FFN). In each cross-attention layer:
\vspace{-1em}
\begin{alignat}{2}\label{eq:cross_attention}
	\mathrm{CrossAttenti}&on(Q_{ca}, &&K_{ca},V_{ca}) = \mathrm{Softmax}(\frac{Q_{ca}K_{ca}^T}{\sqrt{d}}) \cdot V_{ca} \notag \\
		 Q_{ca} & = W_{Q_{ca}}^{(l)} &&\cdot \mathrm{Concat}(\mathbf{X},\mathbf{S}) \notag \\
		 K_{ca} & = W_{K_{ca}}^{(l)} &&\cdot \mathrm{Concat}(\mathbf{Z},\mathbf{V},\mathbf{E}, \mathrm{Emb}(t)) \notag \\
		 V_{ca} & = W_{V_{ca}}^{(l)} &&\cdot \mathrm{Concat}(\mathbf{Z},\mathbf{V},\mathbf{E}, \mathrm{Emb}(t)),
\end{alignat}
where $ l=1,\dots,L $. $ W_{Q_{ca}}^{(l)} $, $ W_{K_{ca}}^{(l)} $ and $ W_{V_{ca}}^{(l)} $ are learnable projection matrices for the $ l $-th cross-attention layer. $ \mathrm{Concat}(\cdot) $ is the concatenation operation, and $ \mathrm{Emb}(t) $ represents the time embedding for the time step $ t $.


\begin{figure*}[t]
	\centering
	\includegraphics[width=1.0\textwidth]{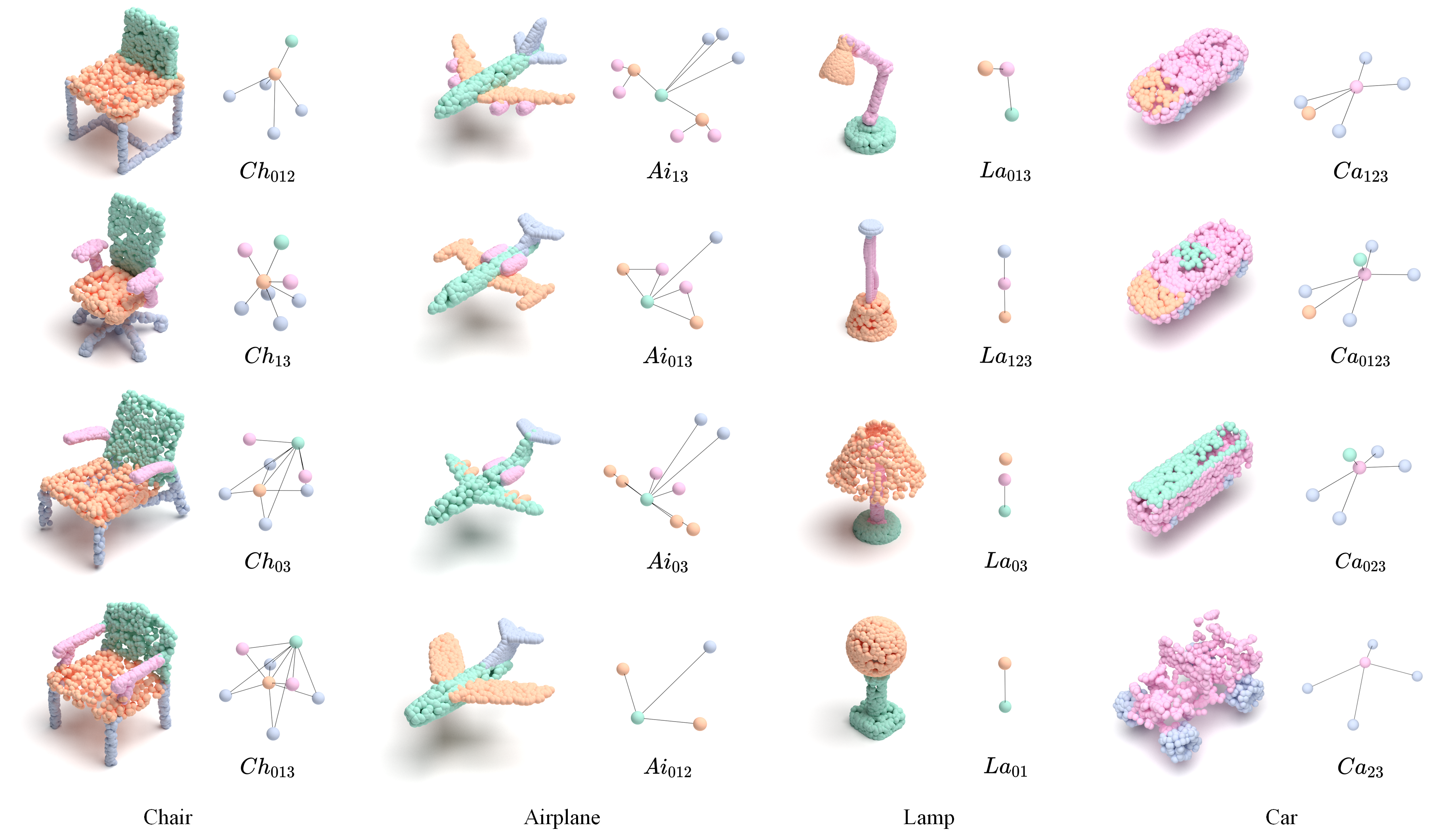}
	\caption{
		3D point cloud shapes~(left column) of the ShapeNet dataset on four categories: Chair, Airplane, Lamp, and Car, with our annotated StructureGraph and their names~(right column). Note that each segmented part in the StructureGraph represents only one graph node, and multiple nodes in the figure are for visualization purposes.
	}
	\label{fig:shapenet_dataset}
\end{figure*}

In the sampling process, as illustrated in Figure~\ref{fig:sampling}, given Gaussian noise as noised point cloud shape $ \hat{\mathbf{X}}^{(T)} \sim \mathcal{N}(0, \mathbf{I}) $ with arbitrary $ \hat{n} $ points, part features $ \hat{\mathbf{Z}} $, semantic segmentation labels $ \hat{\mathbf{S}} $, part existences $ \hat{\mathbf{V}} $, and part adjacency relationships $ \hat{\mathbf{E}} $, reverse diffusion is performed from time step $ T $ to 1. Trained diffusion Transformer $ F_\theta $ predicts the noise and gradually denoises $ \hat{\mathbf{X}}^{(t)} $ to finally generate novel point cloud shapes $ \hat{\mathbf{X}}^{(0)} $ with controllable structure. The sampling process is as follows: 
\begin{align}\label{eq:diff_sampling}
	\hat{\mathbf{X}}^{(t-1)} & = \frac{1}{\sqrt{\alpha_t}}\left(\hat{\mathbf{X}}^{(t)}-\frac{1-\alpha_t}{\sqrt{1-\bar \alpha_t}}F_\theta(\hat{\mathbf{X}}^{(t)},\hat{\mathbf{Z}},\hat{\mathbf{S}},\hat{\mathbf{V}},\hat{\mathbf{E}})\right) \notag \\
	                         & + \sqrt{\frac{1-\bar \alpha_{t-1}}{1-\bar \alpha_t}\beta_t} * \epsilon_t ,
\end{align}
where time step $ t=T,\dots,1 $ and $ \epsilon_t $ is sampled noise at time step $ t $. Note that if the semantic segmentation labels $ \hat{\mathbf{S}} $ are not given, the default semantic segmentation labels can be equally distributed to each point so that each existed segmentation part has the same number of points.

\subsection{Training and Sampling Process}
The overall training loss function $ \mathcal{L} $ for our proposed method is the sum of the loss $ \mathcal{L}_{\mathrm{prior}} $ with loss weight $ \lambda $ for the Prior model $ P_{\psi}\left(\mathbf{Z},\mathbf{V},\mathbf{E}\right) $ to learn the distribution of features $ \mathbf{Z} $ extracted by the StructureGraphNet $ Q_{\phi} (\mathbf{Z}|\mathbf{X},\mathbf{S},\mathbf{V},\mathbf{E}) $ and the loss $ \mathcal{L}_{\mathrm{diff}} $ for the diffusion model $ F_{\theta} $ to predict the noise:
\begin{align}\label{eq:overall_loss}
	\mathcal{L} & = \lambda \mathcal{L}_{\mathrm{prior}} + \mathcal{L}_{\mathrm{diff}} \notag \\
	            & = -\lambda \sum_{j=1}^{m}\Big[\mathbb{E}_{Q_{\phi}\left(z_j|\mathbf{X},\mathbf{S},v_j,e_{j.}\right)}[\mathrm{log}P_{\psi}^{(j)}\left(z_j,v_j,e_{j.}\right)] \notag \\ 
	            & + H\left[Q_{\phi}\left(z_j|\mathbf{X},\mathbf{S},v_j,e_{j.}\right)\right]\Big] \notag \\
				& + \frac{1}{n}\sum_{i=1}^n\mathbb{E}_{\varepsilon,t,\mathbf{Z},\mathbf{S},\mathbf{V},\mathbf{E}}\left[\left\|\varepsilon-F_{\theta}(x_i^{(t)},\mathbf{Z},\mathbf{S},\mathbf{V},\mathbf{E})\right\|_2^2\right] .
\end{align}

To clarify our method, Algorithm \ref{alg:training_process} presents the training process of our approach and Algorithm \ref{alg:sampling_process} presents sampling process.

\section{Experiments}\label{sec:experiments}
This section first provides a detailed overview of our experiments' datasets and evaluation metrics. Then, we present the qualitative and quantitative results of our method on the widely used ShapeNet dataset and StructureNet dataset. Moreover, we compare our method with the results of existing state-of-the-art 3D point cloud generation methods. We also perform some ablation studies to verify the effectiveness of each module in our method. We also show the results of reconstructing surfaces from the generated point clouds. Finally, the implementation details and performance are presented.

\subsection{Experimental Datasets}

\subsubsection{ShapeNet Dataset}
Following prior works~\cite{yang2019pointflow,luo2021dpm,nakayama2023difffacto}, the datasets used in this paper are four categories of the ShapeNet dataset~\cite{chang2015shapenet}: Chair, Airplane, Lamp, and Car, with part segmentation labels~\cite{yi2016shapenetpart}. Each category contains 3053, 2349, 1261, 740 training shapes and 704, 341, 286, 158 test shapes, respectively. On each ShapeNet category, we train our model separately and generate novel point cloud shapes.

\begin{figure*}[t]
	\centering
	\includegraphics[width=1.0\textwidth]{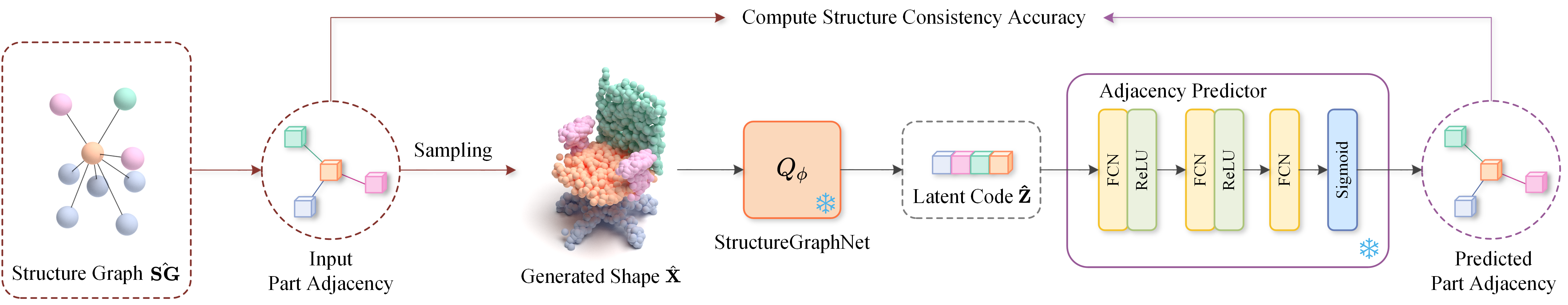}
	\caption{
		Computing process of our proposed Structure Consistency Accuracy between input part adjacency and predicted part adjacency. FCN in the AdjacencyPredictor represents the fully connected neural networks.
	}
	\label{fig:compute_sca}
\end{figure*}

Since the dataset only provides segmentation labels for each shape category, instead of simply computing the minimum distance between two segmented parts to obtain the part adjacency, we manually annotate the adjacency relationships between the segmented parts of the 3D point cloud shapes to construct the StructureGraph. As shown in Figure~\ref{fig:shapenet_dataset}, for each category, the left column is the 3D point cloud shapes with segmentation labels, and the right column is the StructureGraph we annotated along with their names. For example, $ Ch_{13} $ indicates that the $ 1 $-th part~(seat) and the $ 3 $-th part~(armrest) of this chair are adjacent~(we assume that the chair's seat is adjacent to the backrest and leg for brevity). Note that each segmented part in the StructureGraph represents only one graph node, and multiple nodes in the figure are for visualization purposes. We randomly sample 2048 points from each of the point cloud shape and normalize each of them to zero mean and unit variance.

Moreover, we calculate the Cyclomatic Complexity~\cite{mccabe1976complexity} to analyze the complexity of the shape's structure. There are 171 chairs that have reached the maximum complexity of a 4-node complete graph: 10. We also counted that there were a total of 2087 chairs with a complexity of 5 or more. This indicates that the structure of most chairs in the 4-node graph has reached a relatively complex level.

\subsubsection{StructureNet Dataset}
StructureNet~\cite{mo2019structurenet} represents shapes as $n$-ary part hierarchies with adjacency between sibling parts, they compute the minimum distance between two segmented parts to obtain the part adjacency on the PartNet~\cite{mo2019partnet} dataset. The shape structure in the StructureNet dataset is relatively complex.

To adapt the structure of StructureNet dataset to apply in our method, we only use the first level of the root node of the StructureNet data. We trained our method separately on four categories of the StructureNet dataset: Chair, Vase, TrashCan, and Bed. We randomly split the dataset into 85\% training set and 15\% testing set.

\begin{table*}
	\centering
	\caption{Comparing our method with 3D shape generation methods on four categories of the ShapeNet dataset in Shape Generation Metrics. MMD scores and JSD scores are multiplied by $ 10^2 $. COV scores and 1-NNA scores are reported in \%.}
	\label{tab:shapenet_sgm}
	\begin{tabular}{ll|cc|cc|cc|c}
		\toprule
		 &  & \multicolumn{2}{c|}{MMD ($\downarrow$)} & \multicolumn{2}{c|}{COV (\%, $\uparrow$)} & \multicolumn{2}{c|}{1-NNA (\%, $\downarrow$)} & JSD ($\downarrow$) \\ \cmidrule{3-9}
		Shape & Model & CD & EMD & CD & EMD & CD & EMD & - \\
		\midrule
		
		\multirow{6}{*}{Chair}
		& PointFlow~\cite{yang2019pointflow}
			& 9.00     & 35.37     & 22.67     & 25.87      & 93.20     & 96.67     & 5.02 \\
		& DPM~\cite{luo2021dpm}
			& 8.89     & 39.77     & 29.60     & 19.20      & 88.13     & 95.87     & 4.39 \\
		& DiffFacto~\cite{nakayama2023difffacto}
			& 6.28     & 32.83     & 45.37     & \bf 39.13  & 75.48     & \bf 88.54 & 2.75 \\
		& SPAGHETTI~\cite{hertz2022spaghetti}
		    & 10.64	   & 30.37	   & 43.37	   & 32.67	    & 87.74	    & 93.02	    & 3.91     \\
		& SALAD~\cite{koo2023salad}
		    & 10.60	   & \bf 30.21 & 45.07	   & 36.24	    & 85.16	    & 90.08	    & 3.72     \\
		& \bf StrucADT~(Ours)
			& \bf 6.26 & 30.92     & \bf 48.80 & 33.87      & \bf 69.87 & 93.60     & \bf 1.46 \\ 
		\midrule

		\multirow{6}{*}{Airplane}
		& PointFlow~\cite{yang2019pointflow}
			& 3.42     & 23.76     & 35.21     & 34.44     & 87.04     & 90.74     & 2.27 \\
		& DPM~\cite{luo2021dpm}
			& 2.94     & 25.28     & 37.41     & 22.59     & 77.59     & 95.00     & 1.73 \\
		& DiffFacto~\cite{nakayama2023difffacto}
			& 2.81     & 22.08     & 40.20     & 35.34     & 76.23     & \bf 89.07 & 1.62 \\
		& SPAGHETTI~\cite{hertz2022spaghetti}
			& 5.24	   & 25.29	   & 45.56	   & 35.93	   & 89.13	   & 90.78	   & 6.52    \\
		& SALAD~\cite{koo2023salad}
		    & 4.12	   & \bf 21.29 & \bf 47.04 & 33.34	   & 84.97	   & 90.12	   & 6.27   \\
		& \bf StrucADT~(Ours)
			& \bf 2.69 & 23.00     & 42.59     & \bf 37.04 & \bf 74.81 & 94.07     & \bf 1.60 \\ 
		
		\midrule

		\multirow{4}{*}{Lamp} 
		& PointFlow~\cite{yang2019pointflow}
			& 10.82    & 42.09     & 43.87     & 41.29     & 82.26     & 88.06     & 6.74 \\
		& DPM~\cite{luo2021dpm}
			& 11.32    & 40.41     & 43.23     & 38.06     & 80.32     & 93.23     & 4.87 \\
		& DiffFacto~\cite{nakayama2023difffacto}
			& 9.33     & 32.74     & 47.88     & \bf 48.62 & 71.40     & 79.21     & 3.44 \\
		& \bf StrucADT~(Ours)      
			& \bf 8.11 & \bf 31.49 & \bf 49.03 & 45.81     & \bf 61.29 & \bf 68.06 & \bf 2.41 \\ 

		\midrule
		
		\multirow{4}{*}{Car} 
		& PointFlow~\cite{yang2019pointflow}
			& 3.80     & \bf 21.78 & 52.75     & 39.56    & 71.98      & \bf 76.92 & 1.88 \\
		& DPM~\cite{luo2021dpm}
			& 3.54     & 23.37     & 51.65     & 23.08    & 60.44      & 90.66     & 1.86 \\
		& DiffFacto~\cite{nakayama2023difffacto}
			& \bf 3.12 & 22.74     & 55.09     & 43.11    & 62.17      & 80.10     & 1.90 \\
		& \bf StrucADT~(Ours)      
			& 3.34     & 21.93     & \bf 57.14 & \bf 43.96 & \bf 58.24 & 79.67     & \bf 1.84 \\ 
	
		\bottomrule
	\end{tabular}
\end{table*}

\begin{table*}
	\centering
	\caption{Evaluating the Structure Consistency Accuracy of our method on the Chair category. SCA scores are reported in \%.}
	\label{tab:shapenet_sca_chair}
	\begin{tabular}{ll|cccccccc}
		\toprule
		Shape & Model & \tabincell{c}{SCA-$Ch_{012}$} & \tabincell{c}{SCA-$Ch_{03}$} & \tabincell{c}{SCA-$Ch_{13}$} & \tabincell{c}{SCA-$Ch_{23}$} & \tabincell{c}{SCA-$Ch_{013}$} & \tabincell{c}{SCA-$Ch_{023}$} & \tabincell{c}{SCA-$Ch_{123}$} & \tabincell{c}{SCA-$Ch_{0123}$} \\
		\midrule
		
		\multirow{4}{*}{Chair} 
		& PointFlow~\cite{yang2019pointflow}
			& 90.17     & 82.78     & 80.23     & 81.12     & 72.58     & 73.98     & 72.45       & 63.90 \\
		& DPM~\cite{luo2021dpm}
		    & 93.88   	& 82.00     & 81.75     & 83.12     & 73.75     & 73.37     & 73.00       & 63.37 \\
		& DiffFacto~\cite{nakayama2023difffacto}
			& 95.42     & 85.62     & 82.45     & \bf 85.10 & 80.98     & 81.12     & 76.40       & 74.23 \\
		& \bf StrucADT~(Ours)
		    & \bf 97.96	& \bf 91.20 & \bf 85.20 & 82.65     & \bf 87.24 & \bf 83.55 & \bf 77.68   & \bf 77.30 \\ 
	
		\bottomrule
	\end{tabular}
\end{table*}

\subsection{Evaluation Metrics}
We employ two types of metrics to evaluate our proposed structure-controlled 3D point cloud generation method: Shape Generation Metrics and Structure Consistency Accuracy. The Shape Generation Metrics are used to assess the realism and diversity of the generated point cloud shapes while the Structure Consistency Accuracy is utilized to evaluate the consistency between the generated point cloud shapes and the input shape structure.

\emph{Shape Generation Metrics.} Following prior works, we use the Chamfer Distance (CD)~\cite{achlioptas2018rgan} and the Earth Mover's Distance (EMD)~\cite{achlioptas2018rgan} to evaluate the reconstruction quality of the generated point cloud shape. To evaluate the generation quality, we employ the coverage score (COV)~\cite{achlioptas2018rgan}, the minimum matching distance (MMD)~\cite{achlioptas2018rgan}, the Jenson-Shannon divergence (JSD)~\cite{achlioptas2018rgan}, and 1-NN classifier accuracy (1-NNA)~\cite{yang2019pointflow}. MMD is a metric for evaluating the quality of generated point clouds. For each point cloud in the reference set, the distance to its nearest neighbor in the generated set is computed and averaged. COV measures the proportion of point clouds in the reference set that match at least one point cloud in the generated set. For each point cloud in the generated set, its nearest neighbor in the reference set is marked as matched. 1-NNA is used for two-sample tests to assess whether two distributions are identical. JSD is used to calculate the distance between marginal point distributions.

\begin{table*}
	\centering
	\caption{Evaluating the Structure Consistency Accuracy of our method on the Airplane category. SCA scores are reported in \%.}
	\label{tab:shapenet_sca_airplane}
	\begin{tabular}{ll|cccccccc}
		\toprule
		Shape & Model & \tabincell{c}{SCA-$Ai_{012}$} & \tabincell{c}{SCA-$Ai_{03}$} & \tabincell{c}{SCA-$Ai_{13}$} & \tabincell{c}{SCA-$Ai_{23}$} & \tabincell{c}{SCA-$Ai_{013}$} & \tabincell{c}{SCA-$Ai_{023}$} & \tabincell{c}{SCA-$Ai_{123}$} & \tabincell{c}{SCA-$Ai_{0123}$} \\
		\midrule
		
		\multirow{4}{*}{Airplane} 
		& PointFlow~\cite{yang2019pointflow}
			& \bf 85.46 & 78.70     & 77.42     & 75.26     & 71.17     & 69.13     & 69.39     & 61.35 \\
		& DPM~\cite{luo2021dpm}
			& 83.16     & 77.81     & 77.81     & 77.68     & 69.39     & 70.41     & 69.13     & 63.01 \\
		& DiffFacto~\cite{nakayama2023difffacto}
			& 81.22     & 82.16     & 80.73     & \bf 79.18 & 83.90     & 75.02     & 71.74     & 73.95 \\
		& \bf StrucADT~(Ours)    
			& 83.04     & \bf 86.22 & \bf 94.26 & 70.92     & \bf 93.75 & \bf 76.79 & \bf 80.36 & \bf 84.06 \\ 
	
		\bottomrule
	\end{tabular}
\end{table*}

\begin{table*}
	\centering
	\caption{Evaluating the Structure Consistency Accuracy of our method on the Lamp category. SCA scores are reported in \%.}
	\label{tab:shapenet_sca_lamp}
	\begin{tabular}{ll|cccccccc}
		\toprule
		Shape & Model & \tabincell{c}{SCA-$La_{012}$} & \tabincell{c}{SCA-$La_{03}$} & \tabincell{c}{SCA-$La_{13}$} & \tabincell{c}{SCA-$La_{23}$} & \tabincell{c}{SCA-$La_{013}$} & \tabincell{c}{SCA-$La_{023}$} & \tabincell{c}{SCA-$La_{123}$} & \tabincell{c}{SCA-$La_{0123}$} \\
		\midrule
		
		\multirow{4}{*}{Lamp} 
		& PointFlow~\cite{yang2019pointflow}
			& 77.81     & 71.30     & 72.96     & 71.94     & 66.96     & 69.13     & 68.24     & 60.71 \\
		& DPM~\cite{luo2021dpm}
			& 79.08     & 72.83     & 74.11     & 72.19     & 67.09     & 68.75     & 66.71     & 63.01\\
		& DiffFacto~\cite{nakayama2023difffacto}
			& 78.94     & 74.09     & \bf 80.17 & 76.92     & 70.60     & 71.12     & 72.58     & 75.50 \\
		& \bf StrucADT~(Ours)   
			& \bf 81.51 & \bf 77.81 & 78.44     & \bf 78.95 & \bf 78.95 & \bf 77.30 & \bf 76.91 & \bf 79.85 \\ 
	
		\bottomrule
	\end{tabular}
\end{table*}

\begin{table*}
	\centering
	\caption{Evaluating the Structure Consistency Accuracy of our method on the Car category. SCA scores are reported in \%.}
	\label{tab:shapenet_sca_car}
	\begin{tabular}{ll|ccccccc}
		\toprule
		Shape & Model & \tabincell{c}{SCA-$Ca_{03}$} & \tabincell{c}{SCA-$Ca_{13}$} & \tabincell{c}{SCA-$Ca_{23}$} & \tabincell{c}{SCA-$Ca_{013}$} & \tabincell{c}{SCA-$Ca_{023}$} & \tabincell{c}{SCA-$Ca_{123}$} & \tabincell{c}{SCA-$Ca_{0123}$} \\
		\midrule
		
		\multirow{4}{*}{Car} 
		& PointFlow~\cite{yang2019pointflow}
			& 65.43     & 65.82     & 65.18     & 62.37     & 65.94     & 64.27     & 61.35 \\
		& DPM~\cite{luo2021dpm}
			& 65.31     & 68.37     & 67.60     & 64.54     & 63.78     & 67.60     & 63.39 \\
		& DiffFacto~\cite{nakayama2023difffacto}
			& 70.92     & \bf 76.19 & 71.37     & 75.52     & 77.61     & 74.09     & 81.62 \\
		& \bf StrucADT~(Ours)   
			& \bf 76.02 & 74.62     & \bf 76.53 & \bf 86.09 & \bf 88.65 & \bf 86.85 & \bf 98.98 \\ 
	
		\bottomrule
	\end{tabular}
\end{table*}

\emph{Structure Consistency Accuracy.} The core of the structure-controllable generation network proposed in this paper lies in generating point cloud shapes that conform to the input structure. Therefore, in addition to measuring the quality of the generated point cloud shapes, it is also necessary to measure the consistency between the user-input StructureGraph $ \hat{\mathbf{SG}}=\{\hat{\mathbf{S}},\hat{\mathbf{V}},\hat{\mathbf{E}}\} $ and the generated point cloud shapes $ \hat{\mathbf{X}} $. We propose the Structure Consistency Accuracy (SCA) to evaluate this consistency with the following formula:
\begin{align}\label{eq:sca}
	\mathrm{SCA}(\hat{\mathbf{X}},\hat{\mathbf{V}},\hat{\mathbf{E}}) & = \frac{\sum_{\hat v_j \in \hat{\mathbf{V}}} \sum_{\hat e_{j,k} \in \hat{\mathbf{E}}} [q_v * q_e] }{m^2} \notag \\
														q_v & = \mathbb{I}\left[\hat v_j,[N_v(\hat{\mathbf{X}})]_j\right] \notag \\
														q_e & = \mathbb{I}\left[\hat e_{j,k},[N_e(\hat{\mathbf{X}})]_{j,k}\right] ,
\end{align}
where $ j=1,\dots,m $, $ k=1,\dots,m $, and $ N(\cdot) $ represents the neural networks that predict the StructureGraph of the generated point cloud $ \hat{\mathbf{X}} $. Therefore, $ [N_v(\cdot)]_j $ predicts whether the $ j $-th part exists while $ [N_e(\cdot)]_{j,k} $ predicts whether the $ j $-th part is adjacent to the $ k $-th part. As illustrated in Figure~\ref{fig:compute_sca}, the neural networks $ N(\cdot) $ contain frozen StructureGraphNet and AdjacencyPredictor, which are also trained using the point cloud shapes with their corresponding StructureGraph on each category of the ShapeNet dataset. Similarly, $ \hat v_j $ is whether the $ j $-th part exists and $ \hat e_{j,k} $ is whether the $ j $-th part is adjacent to the $ k $-th part in the input StructureGraph. $ \mathbb{I}[a,b] $ is an indicator function that equals 1 if $ a = b $ and 0 otherwise. This metric measures the consistency between the input StructureGraph and the predicted StructureGraph of the generated shape.

\subsection{Experimental Results and Comparisons}
In our work, we evaluate the effectiveness of our proposed method by comparing it with various controllable 3D point cloud generation methods and pre-trained 3D shape generation methods on the ShapeNet dataset. We also evaluate the performance of our method on four categories of the StructureNet dataset.

\begin{figure*}
	\centering
	\includegraphics[width=0.9\textwidth]{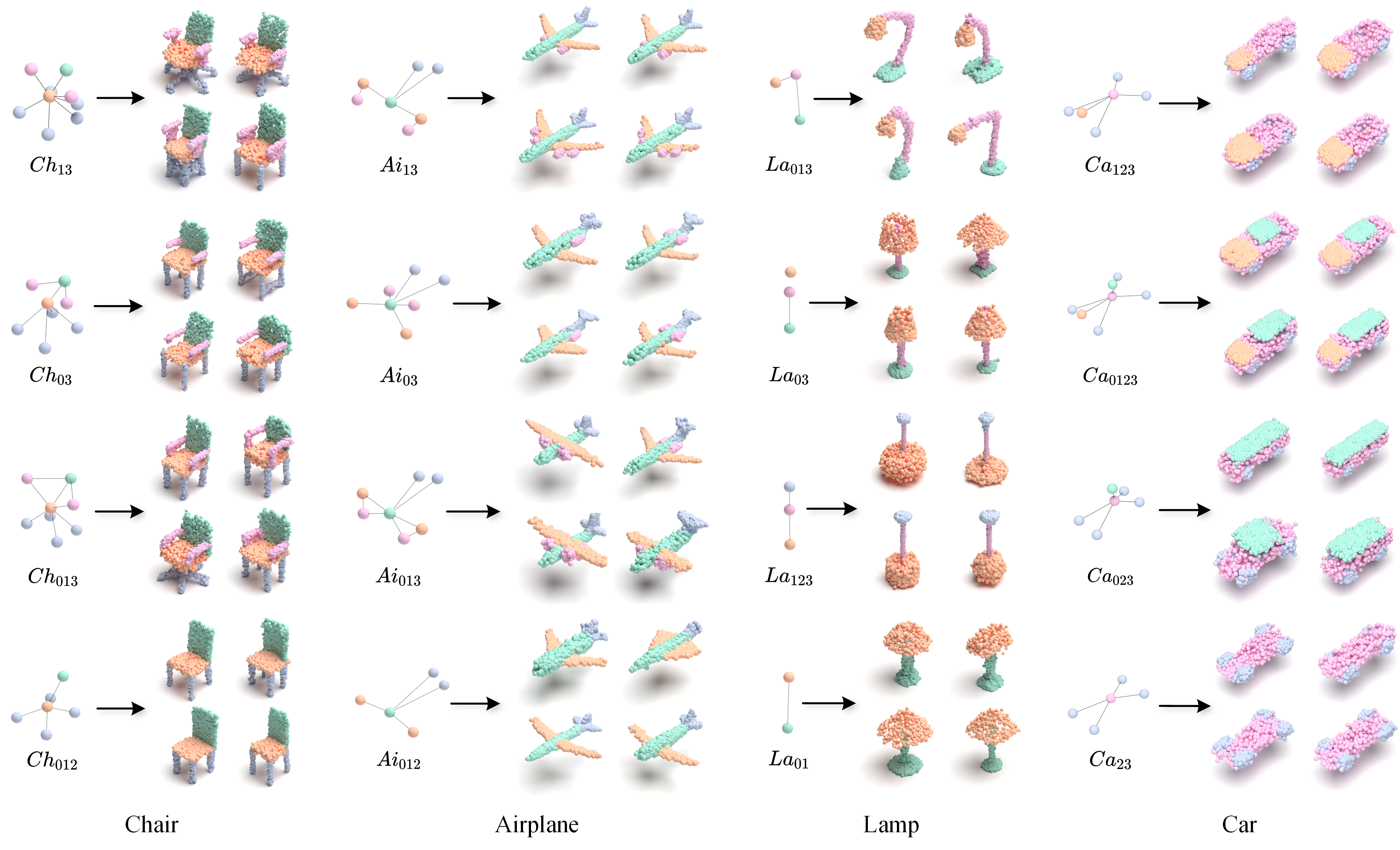}
	\caption{
		Generated 3D point cloud shapes~(right column) controlled by the user-specific StructureGraph~(left column) on four categories of the ShapeNet dataset: Chair, Airplane, Lamp, and Car.
	}
	\label{fig:generated_results}
\end{figure*}

\begin{figure*}
	\centering
	\includegraphics[width=0.9\textwidth]{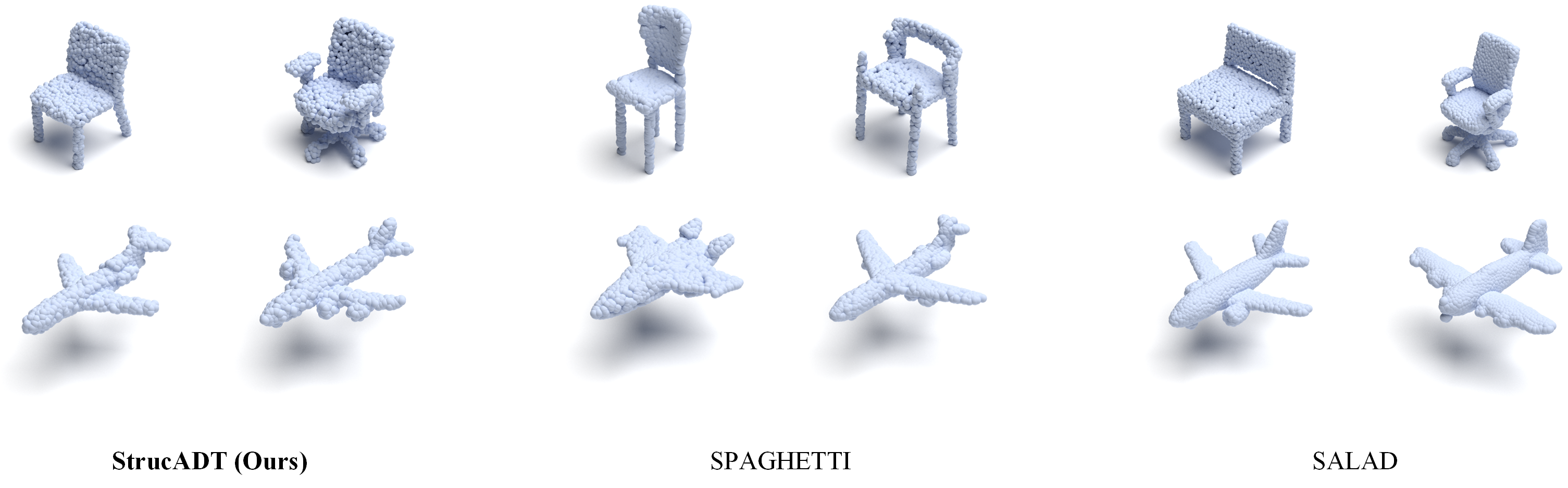}
	\caption{
		Comparison results of generated 3D point cloud shapes between our method (StrucADT) with pre-trained 3D shape generation methods: SPAGHETTI and SALAD on the Chair and Airplane categories of the ShapeNet dataset.
	}
	\label{fig:pretrained_results}
\end{figure*}

		

	

\begin{figure*}
	\centering
	\includegraphics[width=0.9\textwidth]{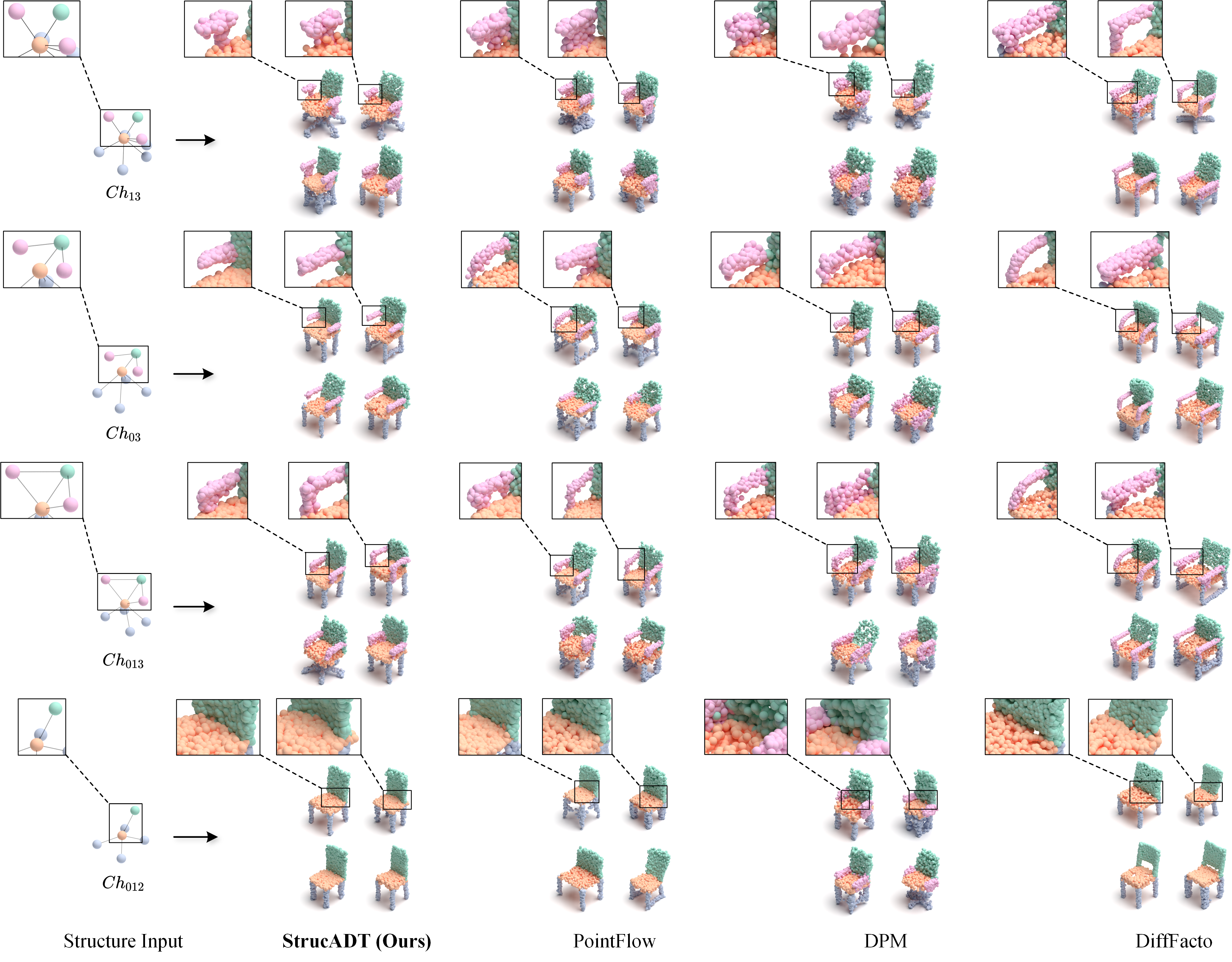}
	\caption{
		Comparison results of generated 3D point cloud shapes controlled by the user-specific StructureGraph on the Chair category of the ShapeNet dataset. The enlarged part of the figure shows the changes in the part adjacency relationships corresponding to different input structures of the chair. Our method can generate shapes consistent with the input structures, while other methods fail.
	}
	\label{fig:compare_results}
\end{figure*}

\subsubsection{ShapeNet Dataset}
Following DiffFacto~\cite{nakayama2023difffacto}, to ensure a fair comparison, we adapt PointFlow~\cite{yang2019pointflow}, DPM~\cite{luo2021dpm}, and DiffFacto~\cite{nakayama2023difffacto} to be controlled by our proposed StructureGraph representation so that we can compare with these methods in Shape Generation Metrics and Structure Consistency Accuracy to evaluate both the generation quality and structure controllability.

\emph{Comparison in Shape Generation Metrics.} 
When comparing in Shape Generation Metrics, we evaluate all the methods on the test shapes of the ShapeNet dataset with our annotated StructureGraph, as illustrated in Figure~\ref{fig:shapenet_dataset}. As shown in Table~\ref{tab:shapenet_sgm}, we compare our method with structure-adapted PointFlow, DPM, and DiffFacto in Shape Generation Metrics on the four categories of the ShapeNet dataset. MMD and JSD scores are multiplied by $ 10^2 $. COV scores and 1-NNA scores are reported in \%. Based on the data in Table~\ref{tab:shapenet_sgm}, our proposed StrucADT attained higher scores in Shape Generation Metrics compared to other methods on each category of the ShapeNet dataset, proving that our method can produce high-quality and varying point clouds controlled by the structure of test shapes. In some metrics of each category, our method is slightly lower than PointFlow and DiffFacto, but our method is generally better than the other three methods on each category. Our experimental results demonstrate that our structure-controlled 3D point cloud generation method outperforms other methods in generating novel and diverse 3D point cloud shapes.

We also compare our method with pre-trained 3D shape generation methods: SPAGHETTI~\cite{hertz2022spaghetti} and SALAD~\cite{koo2023salad}. We used the pre-trained models of the SPAGHETTI and SALAD, and sampled 3D shapes to evaluate the Shape Generation Metrics on the Chair and Airplane categories. As shown in Table~\ref{tab:shapenet_sgm} and Figure~\ref{fig:pretrained_results}, our method outperforms SPAGHETTI and is comparable to SALAD in generating high-quality and diverse point clouds.

\begin{figure*}[t]
	\centering
	\includegraphics[width=0.9\textwidth]{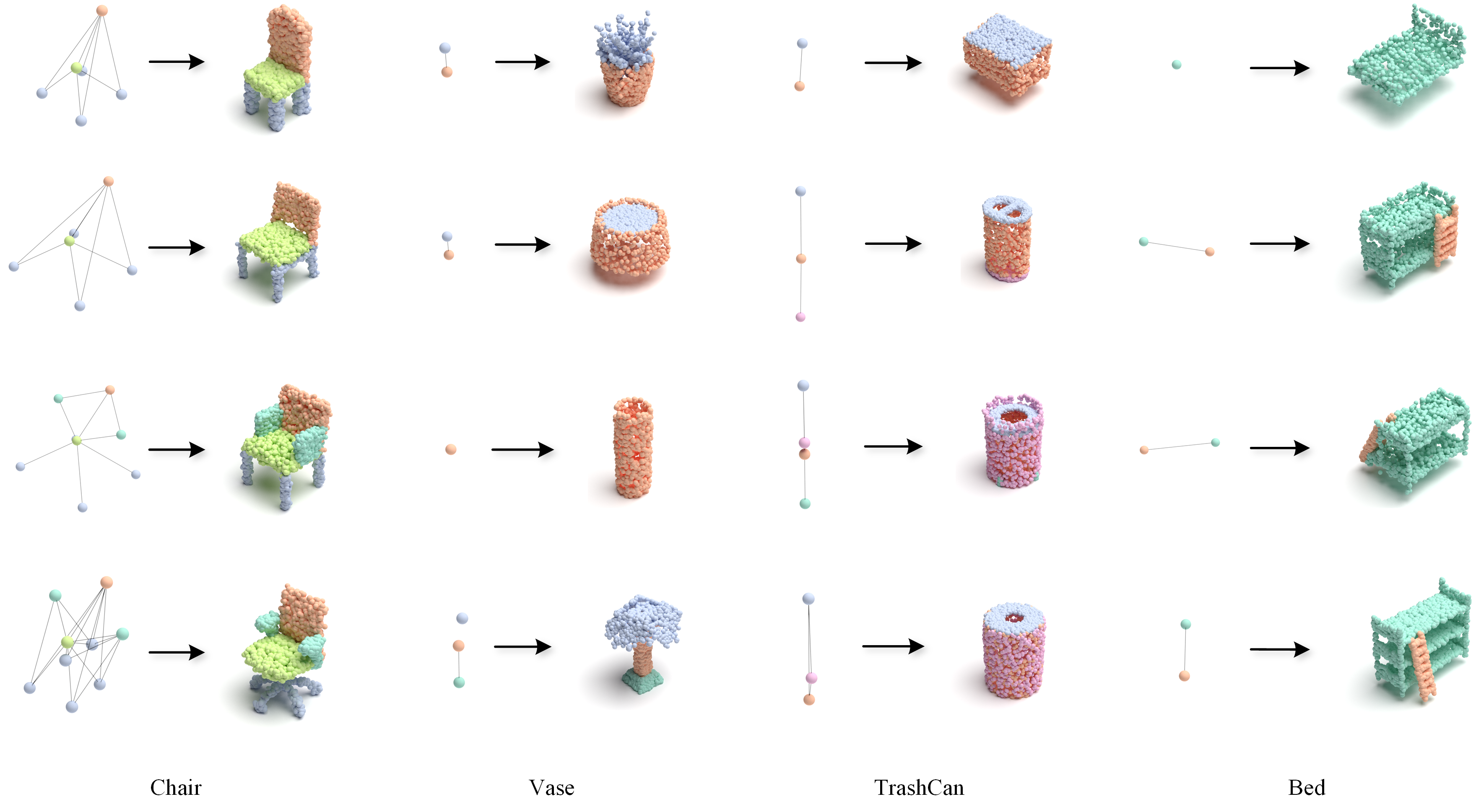}
	\caption{
		Generated 3D point cloud shapes~(right column) controlled by the user-specific StructureGraph~(left column) on four categories of the StructureNet dataset: Chair, Vase, TrashCan, and Bed.
	}
	\label{fig:structurenet_results}
\end{figure*}

\begin{table*}
	\centering
	\caption{Evaluating the performance of our method on four categories of the StructureNet dataset in Shape Generation Metrics. MMD scores and JSD scores are multiplied by $ 10^2 $. COV scores and 1-NNA scores are reported in \%.}
	\label{tab:structure_sgm}
	\begin{tabular}{l|cc|cc|cc|c}
		\toprule
		 & \multicolumn{2}{c|}{MMD ($\downarrow$)} & \multicolumn{2}{c|}{COV (\%, $\uparrow$)} & \multicolumn{2}{c|}{1-NNA (\%, $\downarrow$)} & JSD ($\downarrow$) \\ \cmidrule{2-8}
		Shape & CD & EMD & CD & EMD & CD & EMD & - \\
		\midrule

		Chair    & 5.95  & 31.71 & 40.29 & 34.07 & 75.95 & 92.98 & 1.13 \\
		Vase     & 8.43  & 31.78 & 53.45 & 48.28 & 51.29 & 60.78 & 1.84 \\
		TrashCan & 5.96  & 9.07  & 58.97 & 48.72 & 47.44 & 73.08 & 2.00 \\
		Bed	     & 10.79 & 38.38 & 50.00 & 46.15 & 42.31 & 53.85 & 5.14 \\
	
		\bottomrule
	\end{tabular}
\end{table*}

\emph{Comparison in Structure Consistency Accuracy.} 
When comparing in Structure Consistency Accuracy, the input StructureGraph consists of the user-specific part existences, part adjacency relationships, and the default semantic segmentation labels that give each existing segmentation part the same number of points. We generate 50 shapes for one input StructureGraph on each category of the ShapeNet dataset to compute the averaged Structure Consistency Accuracy. Although our method can generate point clouds with arbitrary points, the generated shapes still contain 2048 points, the same as the input shapes. Figure~\ref{fig:generated_results} displays the generation results of our proposed StrucADT controlled by the user-specific input StructureGraph on the four categories of the ShapeNet dataset. For each category, the left column is the input structures to control point cloud generation, while the right column is the four generated shapes controlled by each input structure. All our generated point cloud shapes are consistent with the input structures on the four categories.

As shown in Table~\ref{tab:shapenet_sca_chair} to Table~\ref{tab:shapenet_sca_car}, we compare our method with structure-adapted PointFlow, DPM, and DiffFacto in Structure Consistency Accuracy on each category of the ShapeNet dataset. SCA scores are reported in \%. We experiment with all the structures occurring on the four categories of the ShapeNet dataset, where the structures with the highest occurrences are listed in the table. For example, SCA-$ Ch_{03} $ represents the Structure Consistency Accuracy of the generated chair with the $ 3 $-th part~(armrests) only adjacent to the $ 0 $-th part~(back), and SCA-$ Ch_{012} $ represents the Structure Consistency Accuracy of the generated chair with no $ 3 $-th part~(armrests). Note that $ Ca_{012} $ does not exist in the Car category because all cars have the $ 3 $-th part~(body). Based on the data in Table~\ref{tab:shapenet_sca_chair} to Table~\ref{tab:shapenet_sca_car}, our proposed StrucADT achieves higher scores in Structure Consistency Accuracy compared to other methods on each category of the ShapeNet dataset. In addition, we find that the Structure Consistency Accuracy decreases as the part adjacency of the input structure increases (e.g., from $Ch_{012}$ to $ Ch_{0123} $) on the Chair category, indicating that more complex structures are more challenging to control the generation of the corresponding 3D point cloud shapes. It is necessary to balance the quality of point cloud generation and structural consistency. Therefore, our method may not achieve the best quality performance of the point clouds generated in some cases, but it can generate point clouds corresponding to the structure input by the user.

Figure~\ref{fig:compare_results} shows a qualitative comparison of our method with the other three methods on the chair category of the ShapeNet dataset. The enlarged part of the figure shows the changes in the part adjacency relationships corresponding to different input structures of the chair. Our method can generate shapes that are consistent with the input structures. Especially when the input structures are $ Ch_{13} $ and $ Ch_{03} $, our method can generate chairs whose armrests are only attached to the seat and whose armrests are only attached to the back, respectively, while other methods fail to generate these shapes. Our experimental results demonstrate that our 3D point cloud generation method can generate shapes controlled by the input structures, outperforming other methods and achieving better structure controllability.

\begin{figure*}[t]
	\centering
	\includegraphics[width=0.7\textwidth]{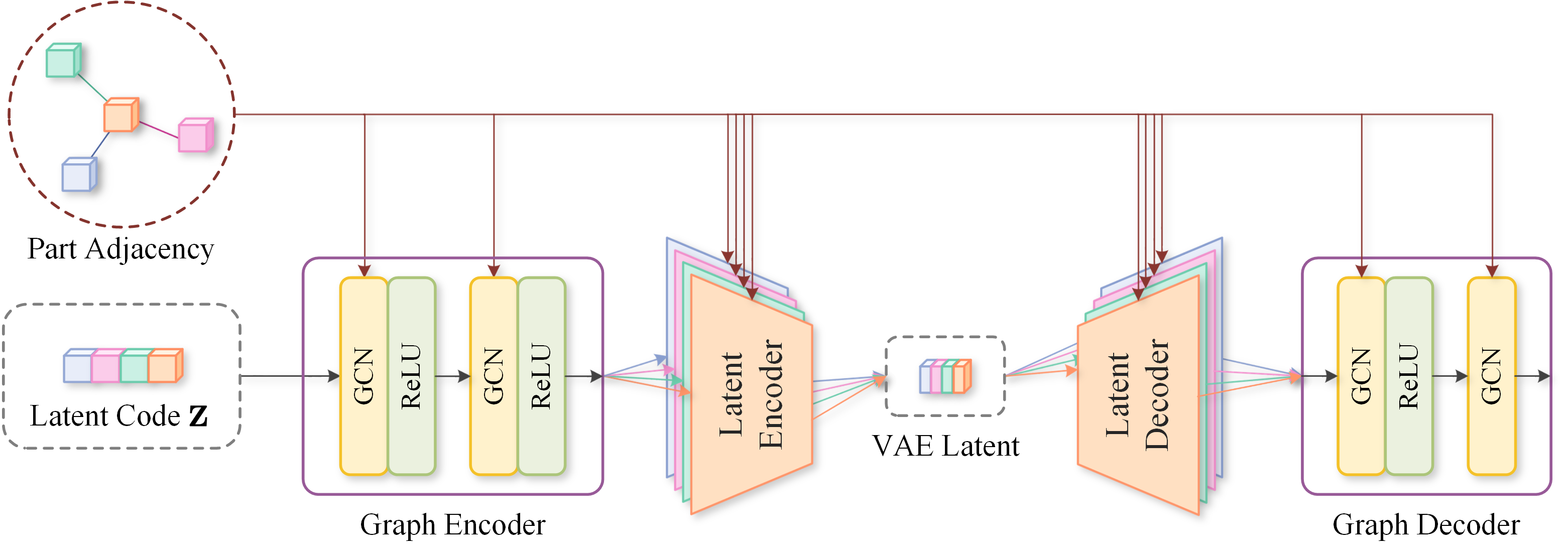}
	\caption{
		Network architecture of GraphCVAE in the ablation experiments of the Prior module. GCN in the Graph Encoder and Graph Decoder represents the graph convolutional neural networks.
	}
	\label{fig:ablation_graphcvae}
\end{figure*}

\begin{figure*}[t]
	\centering
	\includegraphics[width=0.8\textwidth]{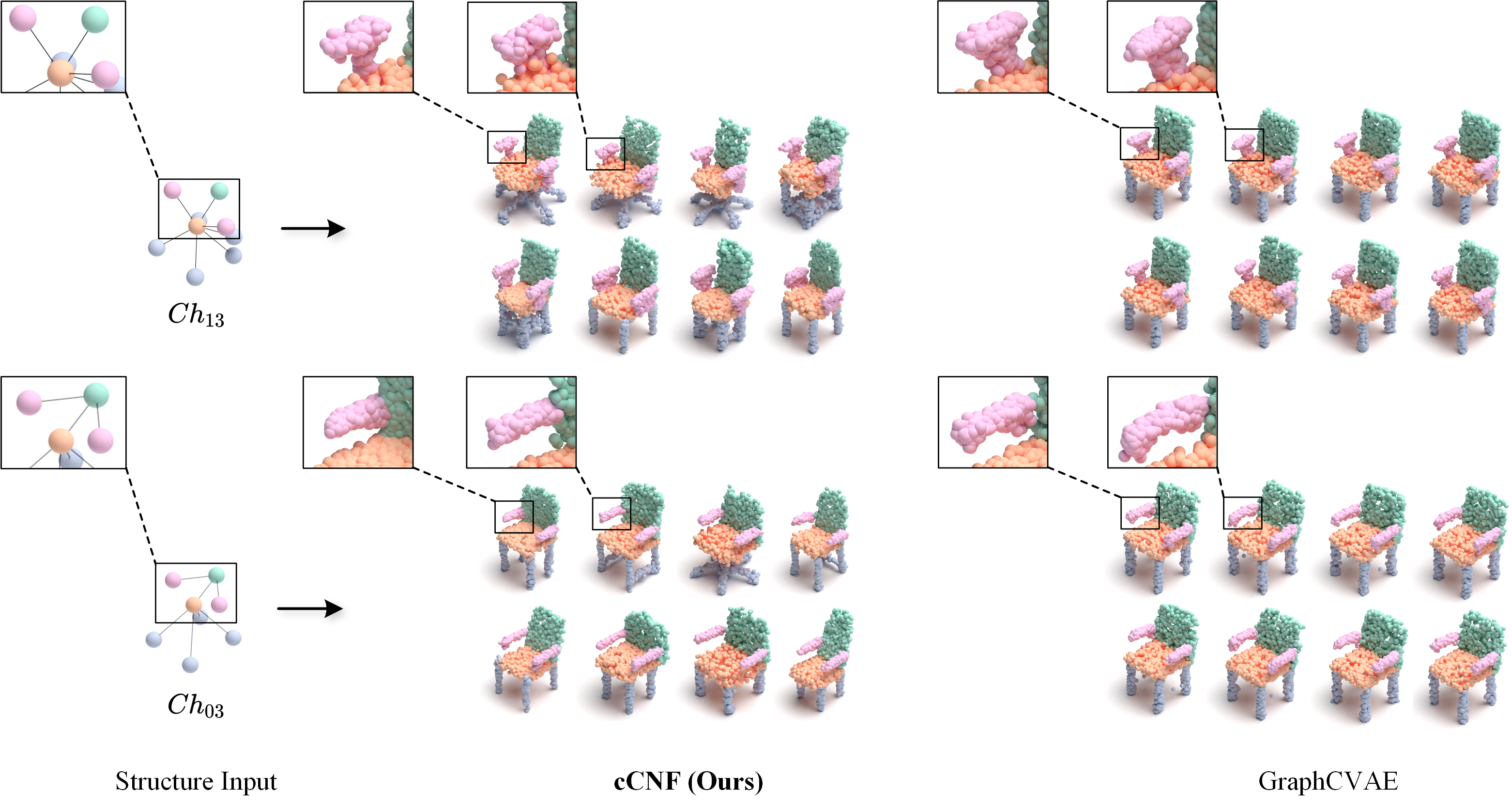}
	\caption{
		Comparison results of ablation experiments on network framework of prior module controlled by the user-specific StructureGraph on chair category. Compared to GraphCVAE, our method generates shapes consistent with the input structures and ensures the quality and diversity of the generated shapes. In contrast, GraphCVAE can only generate similar and repetitive shapes with excessive control.
	}
	\label{fig:ablation_graphcvae_results}
\end{figure*}

\subsubsection{StructureNet Dataset}
We also evaluate the performance of our method on four categories of the StructureNet dataset in Shape Generation Metrics, as shown in Table~\ref{tab:structure_sgm}. Table~\ref{tab:structure_sgm} and Figure~\ref{fig:structurenet_results} indicate that our method can generate high-quality point cloud shapes consistent with the input structure on the relatively complex StructureNet dataset. 

It is worth noting that the StructureGraph encodes only abstract part existences and adjacency relationships, without spatial location or orientation. Hence, the diversity in part layouts seen in the last column of Figure~\ref{fig:structurenet_results} arises solely from the stochastic nature of the diffusion-based generation process. This design allows our model to flexibly generate multiple plausible spatial configurations that are consistent with the same abstract structure.

Moreover, our method naturally supports disconnected nodes in the input StructureGraph, i.e., parts without any adjacent connections. These nodes are encoded individually in the StructureGraphNet without receiving messages from neighbors, and their spatial placement during generation is determined stochastically. For example, in the last Vase instance in Figure~\ref{fig:structurenet_results}, the top part is structurally disconnected from the body, and our model successfully generates it as a separate, non-adjacent component, demonstrating correct handling of isolated parts under structure control.

\subsection{Ablation Study}
In this section, we conduct ablation experiments on the critical modules of our method and evaluate the structure generalization of our method. The other modules remain unchanged when one module is ablated in the experiments.

\subsubsection{StructureGraphNet}
To validate the contributions of our proposed StructureGraphNet~(SGN) to the shape generation's effectiveness, we ablate our StructureGraphNet with the PointNet encoder~\cite{qi2017pointnet}, which only takes in each shape part without part adjacency. Table~\ref{tab:ablation_framework_sgm} shows that our proposed StructureGraphNet outperforms the PointNet encoder in Shape Generation Metrics, and Table~\ref{tab:ablation_framework_sca} displays that our proposed StructureGraphNet achieves higher Structure Consistency Accuracy compared to the PointNet encoder, proving the effectiveness of the StructureGraphNet in extracting structure-aware latent features.

\begin{table*}
	\centering
	\caption{Ablation experiments on the network frameworks of the StructureGraphNet module, Prior module, Diffusion Transformer module and our overall framework using the Chair category in Shape Generation Metrics. SGN is the abbreviation of our proposed StructureGraphNet. DiT is the abbreviation of our Diffusion Transformer module. MMD scores and JSD scores are multiplied by $ 10^2 $. COV scores and 1-NNA scores are reported in \%.}
	\label{tab:ablation_framework_sgm}
	\begin{tabular}{ll|cc|cc|cc|c}
		\toprule
		 &  & \multicolumn{2}{c|}{MMD ($\downarrow$)} & \multicolumn{2}{c|}{COV (\%, $\uparrow$)} & \multicolumn{2}{c|}{1-NNA (\%, $\downarrow$)} & JSD ($\downarrow$) \\ \cmidrule{3-9}
		Shape & Model & CD & EMD & CD & EMD & CD & EMD & - \\
		\midrule
		
		\multirow{2}{*}{Chair} 
		& PointNet~\cite{qi2017pointnet}
			& \bf 6.23 & 32.38     & 48.00     & \bf 36.00 & 71.73     & 96.13     & 2.08 \\
		& \bf SGN~(Ours)      
			& 6.26     & \bf 30.92 & \bf 48.80 & 33.87     & \bf 69.87 & \bf 93.60 & \bf 1.46 \\ 

		\midrule

		\multirow{2}{*}{Chair} 
		& GraphCVAE
			& 10.28    & 36.18     & 29.33     & 25.87     & 93.47     & 98.27     & 8.49 \\
		& \bf cCNF~(Ours)    
			& \bf 6.26 & \bf 30.92 & \bf 48.80 & \bf 33.87 & \bf 69.87 & \bf 93.60 & \bf 1.46 \\ 

		\midrule

		\multirow{3}{*}{Chair} 
		& SGN+PointFlow~\cite{yang2019pointflow}
			& \bf 6.15 & 31.83     & 38.54     & \bf 35.93 & 71.47     & 95.07     & 1.91 \\ 
		& SGN+DPM~\cite{luo2021dpm}
			& 6.51     & 34.53     & 39.73     & 21.07     & 80.53     & 96.13     & 2.39 \\
		& \bf SGN+DiT~(Ours)    
			& 6.26     & \bf 30.92 & \bf 48.80 & 33.87     & \bf 69.87 & \bf 93.60 & \bf 1.46 \\  

		\midrule

		\multirow{3}{*}{Chair} 
		& Two Diffusion Models~\cite{peebles2023dit}
		    & 7.30	   & 33.37	   & 42.40	   & 27.73	   & 81.33	   & 97.20	   & 4.76\\ 
		& Single VAE~\cite{sohn2015cvae}
		    & 9.34	   & 43.75	   & 32.53	   & 20.99	   & 85.74	   & 95.56	   & 4.71 \\
		& \bf StrucADT~(Ours)    
			& \bf 6.26 & \bf 30.92 & \bf 48.80 & \bf 33.87 & \bf 69.87 & \bf 93.60 & \bf 1.46 \\  
	
		\bottomrule
	\end{tabular}
\end{table*}

\begin{table*}
	\centering
	\caption{Ablation experiments on the network frameworks of the StructureGraphNet module, Prior module and Diffusion Transformer module using the Chair category in Structure Consistency Accuracy. SGN is the abbreviation of our proposed StructureGraphNet. DiT is the abbreviation of our Diffusion Transformer module. SCA scores are reported in \%.}
	\label{tab:ablation_framework_sca}
	\begin{tabular}{ll|cccccccc}
		\toprule
		Shape & Model & \tabincell{c}{SCA-$Ch_{012}$} & \tabincell{c}{SCA-$Ch_{03}$} & \tabincell{c}{SCA-$Ch_{13}$} & \tabincell{c}{SCA-$Ch_{23}$} & \tabincell{c}{SCA-$Ch_{013}$} & \tabincell{c}{SCA-$Ch_{023}$} & \tabincell{c}{SCA-$Ch_{123}$} & \tabincell{c}{SCA-$Ch_{0123}$} \\
		\midrule
		
		\multirow{2}{*}{Chair} 
		& PointNet~\cite{qi2017pointnet}
			& 97.83     & \bf 91.33 & 84.21     & \bf 83.80 & 86.27     & 83.41     & 76.95     & 75.81 \\
		& \bf SGN~(Ours)
			& \bf 97.96 & 91.20     & \bf 85.20 & 82.65     & \bf 87.24 & \bf 83.55 & \bf 77.68 & \bf 77.30 \\ 
		
		\midrule

		\multirow{2}{*}{Chair} 
		& GraphCVAE
			& 97.83     & \bf 92.09 & \bf 93.24 & \bf 84.69 & \bf 92.47 & \bf 85.46 & 72.96     & \bf 79.72 \\
		& \bf cCNF~(Ours)
			& \bf 97.96 & 91.20     & 85.20     & 82.65     & 87.24     & 83.55     & \bf 77.68 & 77.30 \\ 

		\midrule

		\multirow{3}{*}{Chair} 
		& SGN+PointFlow~\cite{yang2019pointflow}
			& 97.83     & \bf 91.63 & 85.13     & 82.25     & 86.25     & 82.37     & \bf 79.25 & 76.37 \\
		& SGN+DPM~\cite{luo2021dpm}
			& 96.94     & 88.14     & 85.01     & 79.72     & 85.46     & 79.46     & 78.32     & 76.53 \\
		& \bf SGN+DiT~(Ours)
			& \bf 97.96 & 91.20     & \bf 85.20 & \bf 82.65 & \bf 87.24 & \bf 83.55 & 77.68     & \bf 77.30 \\ 
	
		\bottomrule
	\end{tabular}
\end{table*}

\begin{table*}
	\centering
	\caption{Evaluating the structure generalization of our method on the Chair category of the ShapeNet dataset in Shape Generation Metrics under different Structure Ratios. Structure Ratios refer to the percentage of training samples containing the target structure $Ch_{013}$. MMD scores and JSD scores are multiplied by $ 10^2 $. COV scores and 1-NNA scores are reported in \%.}
	\label{tab:structure_generalization_sgm}
	\begin{tabular}{ll|cc|cc|cc|c}
		\toprule
		 &  & \multicolumn{2}{c|}{MMD ($\downarrow$)} & \multicolumn{2}{c|}{COV (\%, $\uparrow$)} & \multicolumn{2}{c|}{1-NNA (\%, $\downarrow$)} & JSD ($\downarrow$) \\ \cmidrule{3-9}
		Shape & Structure Ratios & CD & EMD & CD & EMD & CD & EMD & - \\
		\midrule
		
		\multirow{6}{*}{Chair} 
		&  0\%          & 7.40	    & 33.25	    & 40.06	    & 32.81	    & 79.81	    & 94.48	    & 2.54 \\
		& 10\%          & 7.39	    & 34.16	    & 41.96	    & 33.22	    & 79.20	    & 94.23	    & 2.91 \\
		& 20\%          & 7.59	    & 32.79	    & 42.13	    & 32.16	    & 78.94	    & 96.06	    & 2.16 \\
		& 30\%          & 7.17	    & 31.72	    & 42.29	    & 34.94	    & 78.38	    & \bf 92.34	& 2.12 \\
		& 50\%          & 6.96	    & 31.84	    & 43.13	    & \bf 35.22	& 77.99	    & 93.51	    & 2.05 \\
		& 85\% (Full)	& \bf 6.26  & \bf 30.92	& \bf 48.80	& 33.87     & \bf 69.87	& 93.60	    & \bf 1.46 \\
	
		\bottomrule
	\end{tabular}
\end{table*}

\subsubsection{cCNF Prior}
We ablate our cCNF Prior with our proposed StructureGraph conditioned VAE~(GraphCVAE) Prior, which encodes the latent code $ \mathbf{Z} $ and decodes it conditioned on the part adjacency with reconstruction loss and latent regularization loss. Figure~\ref{fig:ablation_graphcvae} displays the network architecture of GraphCVAE. As shown in Figure~\ref{fig:ablation_graphcvae_results}, GraphCVAE has more controllability on point cloud generation conditioned on the StructureGraph, but the generated shapes have low diversity and quality. Table~\ref{tab:ablation_framework_sgm} illustrates that our cCNF Prior outperforms the GraphCVAE Prior, achieving higher shape generation scores. Table~\ref{tab:ablation_framework_sca} shows that GraphCVAE Prior has higher Structure Consistency Accuracy than cCNF Prior. Compared to GraphCVAE, our method generates shapes consistent with the input structures and ensures the quality and diversity of the generated shapes. In contrast, GraphCVAE can only generate similar and repetitive shapes with excessive control. Therefore, our proposed cCNF Prior is superior to the GraphCVAE Prior.

\begin{table}
	\centering
	\caption{Evaluating the structure generalization of our method on the Chair category in Structure Consistency Accuracy under different Structure Ratios. Structure Ratios refer to the percentage of training samples containing the target structure $Ch_{013}$. SCA scores are reported in \%.}
	\label{tab:structure_generalization_sca}
	\begin{tabular}{ll|c}
		\toprule
		Shape & Structure Ratios & \tabincell{c}{SCA-$Ch_{013}$} \\
		\midrule
		
		\multirow{6}{*}{Chair} 
		
		& 0\%          & 80.38	     \\
		& 10\%          & 82.62	     \\
		& 20\%          & 84.99	     \\
		& 30\%          & 85.87	     \\
		& 50\%          & 86.49	     \\
		& 85\% (Full)	& \bf 87.24  \\
	
		\bottomrule
	\end{tabular}
\end{table}


\subsubsection{Diffusion Transformer} 
We also ablate our Diffusion Transformer~(DiT) with PointFlow~\cite{yang2019pointflow} and DPM~\cite{luo2021dpm}. SGN+PointFlow utilizes another CNF module controlled by the latent code and part adjacency to learn the distribution of the origin point cloud. SGN+DPM uses DDPM to diffuse point clouds, which is also controlled by the latent code and part adjacency. The denoising networks used in SGN+DPM are a series of fully connected layers with feature concatenation and squashing. Table~\ref{tab:ablation_framework_sgm} and Table~\ref{tab:ablation_framework_sca} show that our SGN+DiT outperforms SGN+PointFlow and SGN+DPM module in both Shape Generation Metrics and Structure Consistency Accuracy, proving the effectiveness of our proposed part adjacency conditioned diffusion Transformer on structure controllable point cloud generation.


\begin{figure*}[t]
	\centering
	\includegraphics[width=1.0\textwidth]{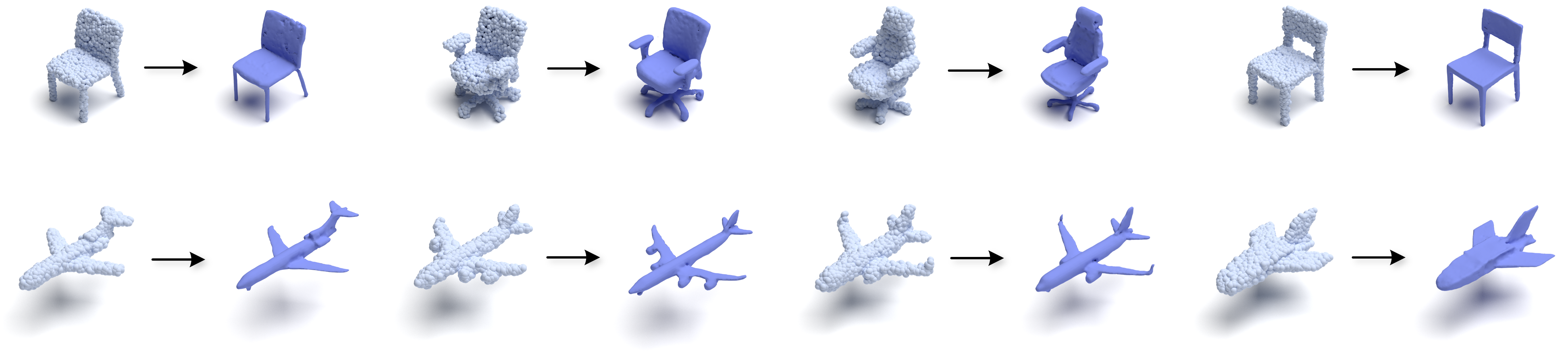}
	\caption{
		The results of reconstructing surfaces from the point clouds generated on the Chair and Airplane categories.
	}
	\label{fig:reconstruction_surface}
\end{figure*}



\begin{figure*}[t]
	\centering
	\includegraphics[width=0.9\textwidth]{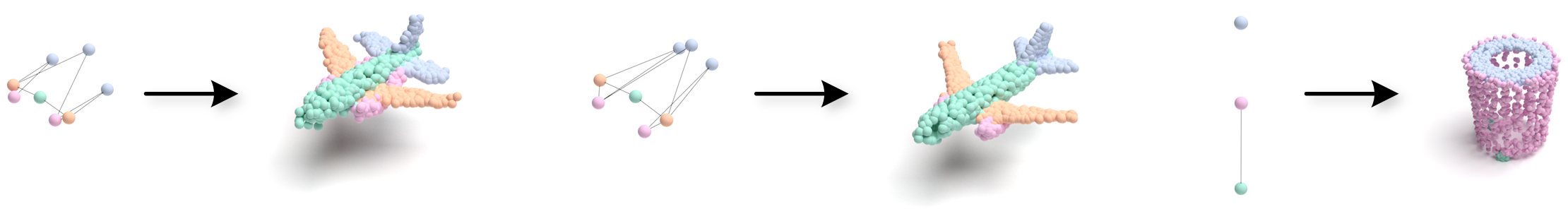}
	\caption{
		When the input structure is very different from the training set, our method has a specific generalization ability, as the input aircraft engine is to be adjacent to both the wing and the tail wing~(left column). Still, it may ignore this input structure, and the generated aircraft engine is only adjacent to the wing~(middle column). Besides, the top of the generated trash can is still adjacent to the body, although the two parts are not connected in the input structure graph~(right column).
	}
	\label{fig:limitation}
\end{figure*}

\subsubsection{Overall Framework} 
We also performed ablation studies on our overall framework. We compare our overall framework with the framework using two diffusion models and the framework using a single VAE. For the framework of two diffusion models~(DiT~\cite{peebles2023dit}), one diffusion model is used to learn latent variables and the other one is used for point cloud generation. For the framework of a single VAE, we utilize the Conditional VAE~\cite{sohn2015cvae}. As shown in Table~\ref{tab:ablation_framework_sgm}, our method outperforms the framework using two diffusion models and the framework using a single VAE.

\subsubsection{Structure Generalization} 

We designed an ablation experiment to evaluate the effectiveness of our method in structure generalization. On the Chair category of the ShapeNet dataset, we regard structure $Ch_{013}$~(the chair's armrest is connected to both the seat and the backrest) as the novel structure and use it as the test set, and other structures as training sets. To assess the generalization ability, we vary the proportion of samples with the structure $Ch_{013}$ included in the training set, which we refer to as the Structure Ratios. We experiment with Structure Ratios of 0\%, 10\%, 20\%, 30\%, 50\%, and 85\%, where 85\% corresponds to the original full training setup. As shown in Table~\ref{tab:structure_generalization_sgm} and Table~\ref{tab:structure_generalization_sca}, when the Structure Ratios is 0\%~(i.e., $Ch_{013}$ is entirely unseen during training), our method is still able to generate shapes that are reasonably consistent with the desired structure, indicating its ability to generalize to novel structural configurations. As more structures $Ch_{013}$ are added to the training set, both the structure consistency accuracy and the quality of the generated point clouds gradually improve. These results demonstrate that our method exhibits a strong capacity for generalizing to unseen or rare structures by leveraging structural priors learned from other configurations.

\subsection{Surface Reconstruction Application}
Figure~\ref{fig:reconstruction_surface} shows the results of reconstructing surfaces from the point clouds generated on the Chair and Airplane categories. We use the pretrained SDF regression model of the Point-E~\cite{nichol2022pointe} to produce meshes from point clouds.

We follow the previous work, DiffFacto~\cite{nakayama2023difffacto}, and randomly downsample the point clouds with more than 2048 points in the point cloud dataset to 2048 points. Therefore, the point clouds generated by random sampling are relatively challenging to reconstruct into meshes. In order to be more suitable for surface reconstruction, we provide the results of the point clouds generated by the dataset with the farthest point sampling, as well as the meshes reconstructed on these point clouds. Figure~\ref{fig:reconstruction_surface} indicates that the point cloud shapes we generated can be well reconstructed as meshes.

\subsection{Implementation Details and Performance}
We implement the proposed algorithm using Python. We use the public pytorch implementation of DDPM as the foundation for our point cloud generation model, which is trained using a batch size of 128 and an initial learning rate of 0.001 for 100,000 iterations.

Our approach is trained and tested on a PC with an Intel Core i7 CPU, 32GB of RAM, and an NVIDIA GeForce GTX 4090 GPU. Our 3D point cloud generation method consists of two main phases, including the training phase and the sampling phase. In the training phase, our algorithm takes about 2 days to train a single category shape for 100,000 iterations. In the sampling phase, generating 50 structure-controlled 3D point cloud shapes takes about 4 seconds.


\section{Limitations and future work}\label{sec:limitation_future_work}
The shape generation method presented in this paper requires the annotation of segmentation semantic labels for 3D point cloud shapes and the connectivity relationships between segmented parts. A potential follow-up work to this paper is self-supervised controllable 3D point cloud shape generation. This approach learns through self-supervision to use the shape itself as the control condition without the need for additional structural information. This could address the issue of annotation being time-consuming and labor-intensive.

We admit that our method relies heavily on the structures that appear in the training set. The structure of the test input in the above ablation study is relatively similar to that in the training set, so our method can have a specific generalization ability. However, when the input structure is very different from the training set, as shown in Figure~\ref{fig:limitation}, our method may ignore the very different input structure and generate shapes based on the learned structure that is most similar to the input structure. We believe that increasing the data in the dataset or fine-tuning on a specific dataset can alleviate this problem, just like Stable Diffusion~\cite{rombach2022ldm} and Large Language Models~\cite{brown2020gpt3}.

Another possible future direction is combining 3D shape generation with text control information to achieve text-controlled 3D point cloud shape generation. While this paper has realized controllable generation based on 3D shape structure, text is a more abstract representation than shape structure. Realizing this would require a more significant amount of training data and fine-grained shape and text annotation pairs.

\section{Conclusion}\label{sec:conclusion}
In order to address the challenge of lacking control in 3D point cloud generation, this paper leverages the inherent structure of 3D shapes. We control the generation of corresponding point cloud shapes for shapes within the same category by inputting different shape structures. We manually annotate the adjacency relationships between segmented parts of each shape, forming the StructureGraph representation. In this graph, nodes represent segmented parts of the point cloud, while edges denote the connectivity relationships between these segmented parts. Utilizing this StructureGraph representation, we propose StrucADT, a novel structure-controllable point cloud generation model built upon the part adjacency conditioned diffusion Transformer model. The StructureGraphNet module in StrucADT extracts structure-aware latent features, whose distributions are then learned by the cCNF module controlled by part adjacency, and both the latent features and part adjacency are incorporated into the Diffusion Transformer module as conditional context to produce structure-controlled point cloud shapes. Experimental results indicate that our structure-controllable 3D point cloud generation method achieves state-of-the-art performance on the ShapeNet dataset, generating high-quality and diverse point cloud shapes while allowing users to control the generation of corresponding point cloud shapes based on the input shape structures.

\section*{Acknowledgments}\label{sec:acknowledgments}
This work is supported by the National Natural Science Foundation of China (62172356, 61872321, 62272277), Zhejiang Provincial Natural Science Foundation of China (LZ25F020012), the Ningbo Major Special Projects of the ``Science and Technology Innovation 2025'' (2020Z005, 2020Z007, 2021Z012, 2025Z030).

\ifCLASSOPTIONcaptionsoff
	\newpage
\fi

\bibliographystyle{IEEEtran}
\bibliography{mybibfile}

\begin{IEEEbiography}[{\includegraphics[width=1in,height=1.25in,clip,keepaspectratio]{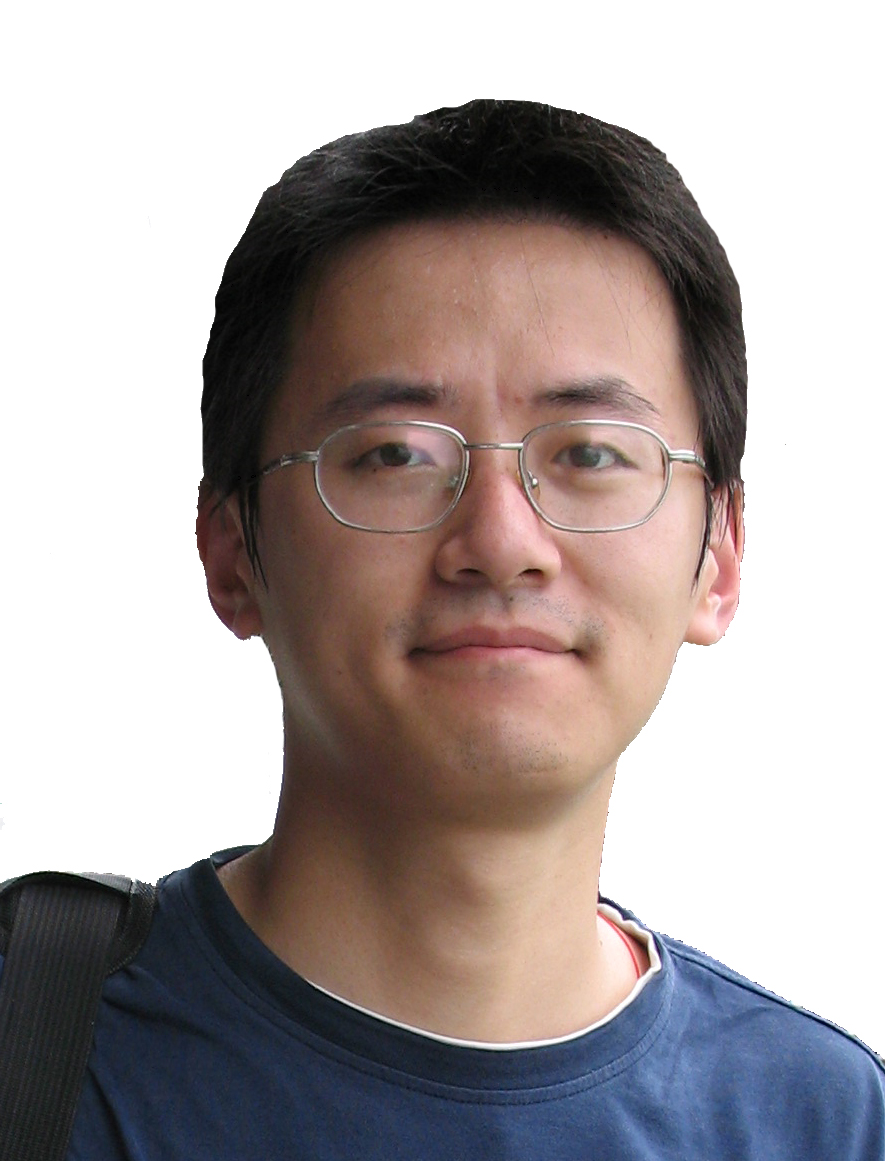}}]{Zhenyu Shu}
	earned his PhD degree in 2010 at Zhejiang University, China. He is now working as a full professor at NingboTech University. His research interests include image processing, computer graphics, digital geometry processing, and machine learning. He has published over 40 papers in international conferences or journals.
\end{IEEEbiography}

\begin{IEEEbiography}[{\includegraphics[width=1in,height=1.25in,clip,keepaspectratio]{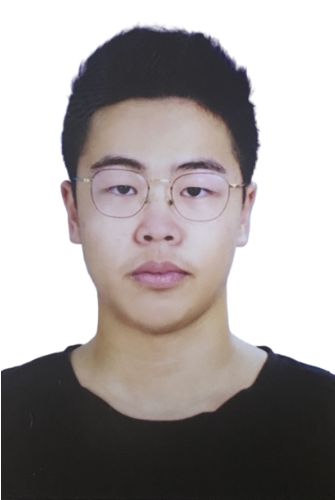}}]{Jiajun Shen}
	is a graduate student of the School of Software Technology at Zhejiang University. His research interests include computer graphics, geometric processing and computer vision.
\end{IEEEbiography}

\begin{IEEEbiography}[{\includegraphics[width=1in,height=1.25in,clip,keepaspectratio]{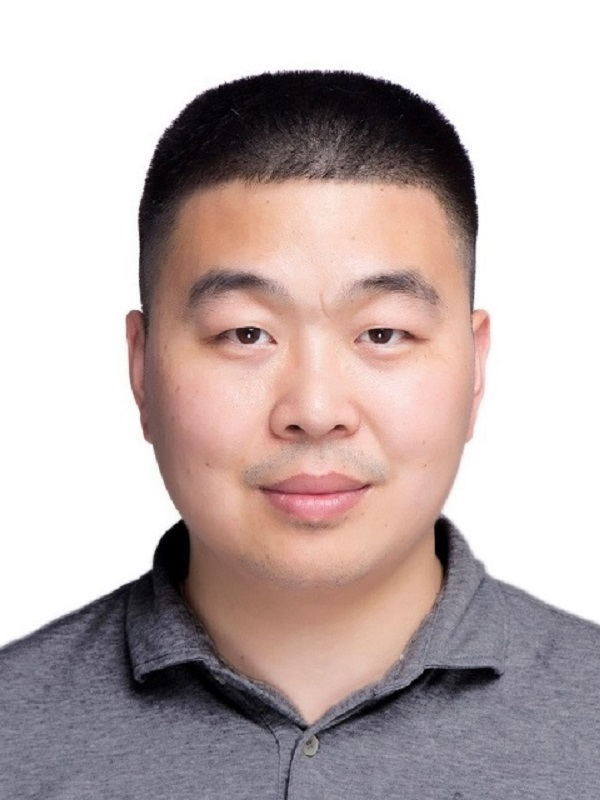}}]{Zhonggui Chen} 
	received BSc. and Ph.D. degrees in applied mathematics from Zhejiang University,
	in 2004 and 2009, respectively. He is working as a full professor in the School of Informatics, Xiamen University. His research interests include computer graphics and computational geometry.
\end{IEEEbiography}

\begin{IEEEbiography}[{\includegraphics[width=1in,height=1.25in,clip,keepaspectratio]{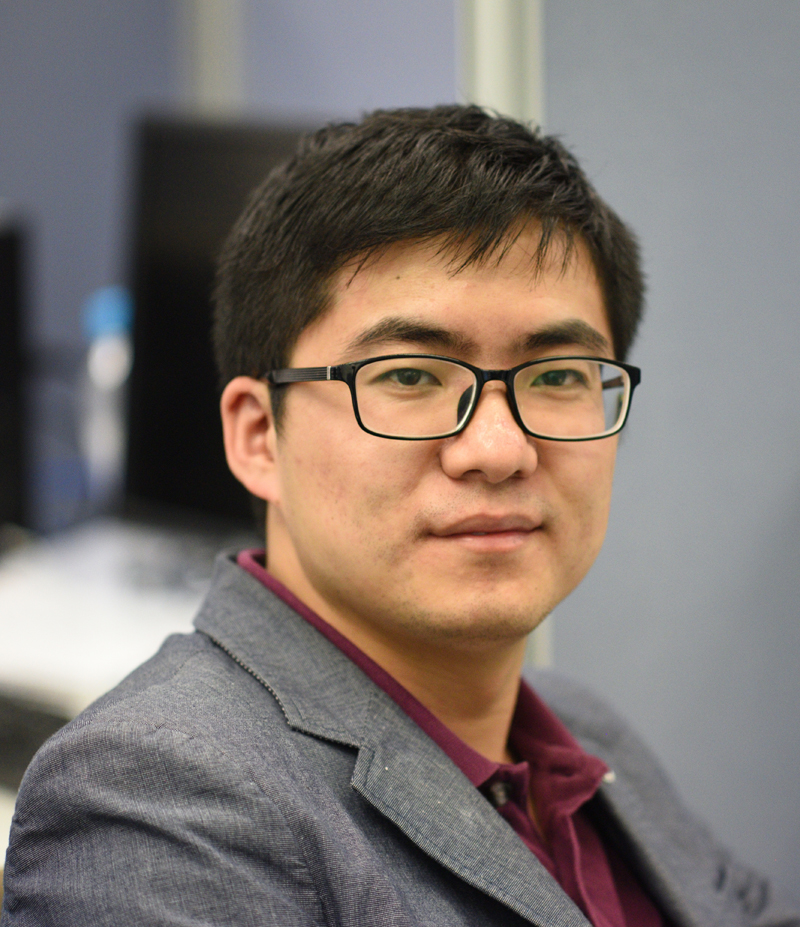}}]{Xiaoguang Han}
	is now an Assistant Professor and President Young Scholar of the Chinese University of Hong Kong (Shenzhen) and the Future Intelligence Network Research Institute. He received his PhD degree from the University of Hong Kong in 2017. His research interests include computer vision and computer graphics. He has published nearly 50 papers in well-known international journals and conferences, including top conferences and journals SIGGRAPH (Asia), IEEE TVCG, CVPR, ICCV, ECCV, NeurIPS, ACM TOG, etc. He currently serves as an associate editor of IEEE Transactions on Visualization and Computer Graphics.
\end{IEEEbiography}

\begin{IEEEbiography}[{\includegraphics[width=1in,height=1.25in,clip,keepaspectratio]{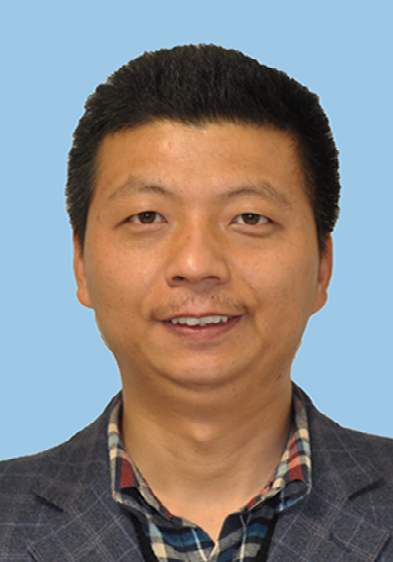}}]{Shiqing Xin}
	is a full professor at the School of Computer Science and Technology at Shandong University. He received his PhD degree in applied mathematics at Zhejiang University in 2009. His research interests include image processing, computer graphics, computational geometry, and 3D printing.
\end{IEEEbiography}


\vfill







\end{document}